\documentclass[12pt,peerreview,draftclsnofoot]{IEEEtran}

\usepackage{amsmath}
\usepackage{cite}
\usepackage{latexsym}
\usepackage{enumerate}
\usepackage{graphics}
\usepackage{graphicx}
\usepackage{algorithm}
\usepackage{algorithmic}
\usepackage{amsfonts}
\usepackage{amssymb}

\newtheorem{theorem}{Theorem}
\newtheorem{lemma}{Lemma}
\newtheorem{proposition}{Proposition}
\newtheorem{definition}{Definition}

\newtheorem{remark}{Remark}

\newtheorem{corollary}{Corollary}

\newenvironment{myalgo}{\begin{algorithmic}}{\end{algorithmic}}
\newenvironment{myalgorithm}{\begin{algorithm}}{\end{algorithm}}

\newenvironment{tit}{
\title{The Delay-Limited Capacity Region of OFDM Broadcast Channels$^*$ 
  \thanks{$^*$ The material in this paper was in part presented at 
    IEEE Information Theory Workshop, Punta del Este, March 2006, 
    IEEE SPAWC, Cannes, July 2006 and 
    44th Allerton Conference, Monticello, 2006}
}
\author{Gerhard Wunder,~\IEEEmembership{Member,~IEEE}, and
  Thomas Michel,~ \IEEEmembership{Student Member,~IEEE}$^\dagger$
  \thanks{$^\dagger$ The authors are with the Fraunhofer German-Sino Mobile Communications Lab,
    Heinrich-Hertz-Institut, Einstein-Ufer 37, D-10587
    Berlin, Germany. 
    Tel/Fax: +493031002-872/-863; Email: \{wunder,michel\}@hhi.fhg.de}
  }
  
  \maketitle

\begin{abstract}
   In this work, the delay limited capacity (DLC) of orthogonal
   frequency division multiplexing (OFDM) systems is investigated. The
   analysis is organized into two parts. In the first part, the impact
   of system parameters on the OFDM DLC is analyzed in a general
   setting. The main results are that under weak assumptions the
   maximum achievable single user DLC is almost independent of the
   distribution of the path attenuations in the low signal-to-noise
   (SNR) region but depends strongly on the delay spread. In the high
   SNR region the roles are exchanged. Here, the impact of delay
   spread is negligible while the impact of the distribution becomes
   dominant. The relevant asymptotic quantities are derived without
   employing simplifying assumptions on the OFDM correlation
   structure. Moreover, for both cases it is shown that the DLC is
   maximized if the total channel energy is uniformly spread, i.e. the
   power delay profile is uniform. It is worth pointing out that since
   universal bounds are obtained the results can also be used for
   other classes of parallel channels with block fading
   characteristic. The second part extends the setting to the
   broadcast channel and studies the corresponding OFDM DLC BC region.
   An algorithm for computing the OFDM BC DLC region is presented. To
   derive simple but smart resource allocation strategies, the
   principle of rate water-filling employing order statistics is
   introduced. This yields analytical lower bounds on the OFDM DLC
   region based on orthogonal frequency division multiple access
   (OFDMA) and ordinal channel state information (CSI).  Finally, the
   schemes are compared to an algorithm using full CSI.

\end{abstract}

\newpage\begin{keywords} delay limited capacity, orthogonal frequency
   division multiplexing (OFDM), broadcast channel, power allocation,
   frequency division multiple access (FDMA), subcarrier assignment
\end{keywords}

}

\begin{document}
\tit

\section{Introduction}

Multiple degrees of freedom in fading channels allow reliable communication in
each fading state under a long term power constraint. This is due to the
possibility of recovering the information from several independently faded
copies of the transmitted signal. The rate achievable in each fading state is
called \emph{zero outage capacity} or alternatively \emph{delay limited
capacity }(DLC) \cite{hanly_98_inf}. Not only multiple input multiple output
(MIMO) channels but also frequency selective multipath channels offer multiple
degrees of freedom. This is in contrast to single antenna Rayleigh flat fading
channels, where a DLC exists only if zero is not in the support of the fading
distribution \cite{BigProSha:1998}. Since the DLC does not involve a decoding
delay over multiple fading blocks if the variation of the fading process is
slow enough, it can be considered as an appropriate limit for delay sensitive
services, which become more and more important recently. Unlike the DLC, the
traditional ergodic capacity strongly depends on the correlation structure of
the fading process and generally implies an infinite decoding delay.

This work investigates the DLC of frequency selective multipath channels in
the context of an orthogonal frequency division multiplexing (OFDM) broadcast
(BC) channel. OFDM can be considered as a special case of parallel fading
channels with correlated fading process. Pioneering work on this topic was
carried out in \cite{caire_99_inf} where the general single user outage
capacity was investigated of which the DLC is a special case. The optimal
power control law is derived which is such that bad channels below some
threshold are simply switched off. In \cite{biglieri_01_inf} the limiting
performance in the high signal-to-noise (SNR) regime of the DLC of parallel
fading channels with multiple antennas was characterized assuming that the
fading distribution is continuous. In \cite{jorswieck_05_itg} the impact of
spatial correlations was studied. Further work was carried out in
\cite{LiGol:2001b,HupBos:2005}. Unfortunately, these results do not carry over
to the OFDM case: since the subcarriers are highly correlated due to
oversampling of the channel in the frequency domain the fading distribution is
commonly degenerated which significantly complicates the analysis. This
particularly affects the critical impact of the delay spread and the number of
subcarriers. Hence, the behavior of the OFDM DLC is not clear yet.

Our main contributions are the following: First, we analyze the impact of
system parameters such as delay spread, power delay profile (or the multipath
intensity profile) and the fading distribution for the single user DLC in a
general setting. We focus on two cases: the behaviour at high SNR and at low
SNR. Such approach has been frequently used in the analysis of channel
capacity even if the capacity itself is not completely known
\cite{Ver:1990,Ver:2002}. The low SNR regime is characterized by its first and
second order expansion. It is shown that to become first order optimal it is
sufficient to serve only one of the best subcarriers regardless of the fading
distribution. The corresponding limit is almost independent of the fading
distribution. The second order limit is also calculated and depends generally
on the number of supported subcarriers, i.e. when the fading distribution
contains point masses. The quantities are shown to exist for a large class of
fading distributions and are explicitly calculated in terms of the delay
spread for the Rayleigh fading case. For the high SNR regime several universal
bounds are calculated. These bounds culminate in a convergence theorem that
generally characterizes the high SNR behaviour under very weak assumptions.
Most important, there will be no need for concepts like almost sure
convergence of the empirical distributions etc. (as used in
\cite{biglieri_01_inf}) and one approaches ergodic capacity relatively fast
(the difference decreases with order $\log^{-1}\left(  L\right)  $ in a
channel with $L$ uniform taps and Rayleigh fading) even if the fading gains
are \emph{not independent}. The corresponding convergence processes are
characterized. It is worth pointing out that these results not only hold for
the OFDM case but also for other classes of parallel block fading channels.
Finally, we provide a convergence result with respect to ergodic capacity.

In the second part we focus on the broadcast scenario, complicating the
analysis significantly. To begin, we present an algorithm which is capable of
evaluating the OFDM BC DLC region up to any finite accuracy. This is a
challenging problem, since for each fading state the minimum sum power
supporting a set of rates has to be found \cite{michel_05_allerton}. To get a
guideline for algorithm design, we subsequently derive lower bounds on the
single user OFDM DLC based on rate water-filling and order statistics. These
single user bounds are the point of origin for the development of simple
analytical lower bounds on the OFDM DLC region based on orthogonal frequency
division multiple access (OFDMA). The involved use of order statistics has a
positive impact on the feedback protocol. In the low SNR regime, nearly the
entire OFDM BC DLC region and in the high SNR regime a significant part of it
can be achieved with these schemes without any form of time-sharing. Further,
a practical OFDMA algorithm based on rate water-filling assuming perfect CSI
is introduced. This scheme outperforms the bounds based on partial CSI and
might serve as a benchmark for other OFDMA minimum sum power algorithms. The
results are illustrated by simulations.

The remainder of this paper is organized as follows: Section \ref{sec:model}
introduces the OFDM system model. In Section \ref{sec:su} and \ref{sec:impact}%
, the behavior at low and high SNR is studied in detail for the single user
case. Section \ref{sec:mu} contains the characterization and computation of
the OFDM broadcast channel DLC region. Subsequently, in Section
\ref{sec:ofdma} lower bounds on the OFDM BC DLC region are derived. We
conclude with some final remarks final Section \ref{sec:conclusions}.

\subsection{Notation}

All terms will be arranged in boldface vectors (where $m$ refers to users, $k$
to subcarriers, $l$ to path delays as a guideline) and the corresponding
indices will be omitted if there is no ambiguity. Common vector norms (such as
$\left\|  \cdot\right\|  _{1}$ for the $l_{1}$-norm) will be employed. The
expression $z\sim\mathcal{CN}\left(  0,1\right)  $ means that the random
variable $z=x+jy$ is complex Gaussian distributed, i.e. the real and imaginary
parts are independently Gaussian distributed with zero mean and variance 1/2:
$x,y\sim\mathcal{N}\left(  0,1/2\right)  $. The expectation operator (e.g.
with respect to the fading process) will be denoted as $\mathbb{E}$
(respectively $\mathbb{E}_{\mathbf{h}}$). $\Pr(A)$ denotes the probability of
an event $A$. We write $f\left(  x\right)  \sim g\left(  x\right)  $ if
$f(x)/g(x)\rightarrow1$ if $x\rightarrow0$ (or $x\rightarrow\infty$) and all
logarithms are to the base $e$ unless explicitely defined in a different manner.

%
%
%
%
%
%
%
%
%
%
%
%
%
%
%
%
%
%
%
%
%
%
%
%
%
%
%
%
%

\section{System Model}

\label{sec:model}

Assume an OFDM broadcast channel with $M$ users from the set $\mathcal{M}%
:=\left\{  1,\ldots,M\right\}  $ and $K$ subcarriers from the set
$\mathcal{K}:=\left\{  1,\ldots,K\right\}  $. The sampled frequency response
of each user is by means of Fast Fourier Transform (FFT) given by
\begin{equation}
\tilde{h}_{m,k}=\sum_{l=1}^{L}\tilde{c}_{m,l}\,e^{-\frac{2\pi j\left(
l-1\right)  \left(  k-1\right)  }{K}},\quad k\in\mathcal{K}
\label{eqn:channel}%
\end{equation}
where $L\leq K$ is the delay spread and $\tilde{c}_{m,l}$ are the complex path
gains which are i.i.d. according to $\tilde{c}_{m,l}\sim\mathcal{CN}%
(0,\sigma_{m,l})$. The vector $\boldsymbol{\sigma}_{m}=[\sigma_{m,1}%
,...,\sigma_{m,K}]^{T}$ is called the power delay profile (PDP) of the $m$th
user's channel. The variances $\sigma_{m,l}$ are assumed to be strictly
positive for all $m$ and $l$ and the channel energy is normalized $\left\|
\boldsymbol
{\sigma}_{m}\right\|  _{1}=1$ for all users $m\in\mathcal{M}$. We say that the
channel of user $m$ has a uniform PDP if $\sigma_{1,m}=\ldots=\sigma_{L,m}$
and a non-uniform PDP otherwise. Note that in practice the PDP is typically
non-uniform. Furthermore, the channel gains are not spread over the entire
frequency band. Then, our results hold approximately and serve also as a
performance limit. The channel (path) gains are defined as $h_{m,k}%
:=|\tilde{h}_{m,k}|^{2}$ (respectively $c_{m,l}:=|\tilde{c}_{m,l}|^{2}$). The
distribution of the channel gains is called the (joint) fading distribution.
In case of complex Gaussian distributed path gains, i.e. $\tilde{c}_{k}%
\sim\mathcal{CN}\left(  0,1/L\right)  $ the channel gains follow a exponential
distribution with $\Pr\left(  h_{k}>x\right)  =e^{-x}$. This case corresponds
to Rayleigh fading.

Let $x_{m,k}$ with power $p_{m,k}$ be the signal transmitted from the base
station to user $m$ on carrier $k$ and let $\mathbf{x}_{m}=[x_{m,1}%
,...,x_{m,K}]^{T},\mathbf{p}_{m}=[p_{m,1},...,p_{m,K}]^{T}$ the stacked vector
of transmit signals and transmit powers for user $m$, respectively. Further
assume that the system is limited by a sum power constraint ${\mathbb{E}%
}\left(  \sum_{m=1}^{M}\left\|  \mathbf{x}_{m}\right\|  _{2}^{2}\right)  \leq
KP^{\ast}$ where $P^{\ast}$ is the power budget. Then the system equation on
each subcarrier $k$ can be written as
\[
y_{m,k}=\tilde{h}_{m,k}\sum_{i\in\mathcal{M}}x_{i,k}+n_{m,k},\quad
k\in\mathcal{K},
\]
where $y_{m,k}$ is the signal received by user $m$ on subcarrier $k$, and
$n_{m,k}$ represents (without loss of generality) normalized noise
$n_{m,k}\sim\mathcal{CN}\left(  0,1\right)  $. Let us define a decoding order
$\pi$, such that user $\pi\left(  M\right)  $ is decoded first, followed by
user $\pi\left(  M-1\right)  $ and so on. Assuming ideal superposition coding
at the transmitter and successive interference cancellation with the decoding
order $\pi$ at the receivers, the rate of user $\pi(m)$ is then given by
\begin{equation}
\tilde{R}_{\pi\left(  m\right)  }=\frac{1}{K}\sum_{k=1}^{K}\log\left(
1+\frac{h_{\pi\left(  m\right)  ,k}p_{\pi\left(  m\right)  ,k}}{1+h_{\pi
\left(  m\right)  ,k}\sum_{n<m}p_{\pi\left(  n\right)  ,k}}\right)
\quad\text{in [nats/Hz/s]}. \label{rate_bc}%
\end{equation}
For an instantaneous channel realization $\mathbf{h}=[h_{1,1},...,h_{1,K}%
,h_{2,1},...,h_{M,K}]^{T}$ (respectively $\mathbf{c}=[c_{1,1},...,c_{1,L}%
,c_{2,1},...,c_{M,L}]^{T}$ is the vector of path gains) the OFDM broadcast
channel capacity region is given by the union over all power allocations
fulfilling the sum power constraint $P^{\ast}$ and over the set of all
possible permutations $\Pi$:
\[
\mathcal{C}_{\mathrm{BC}}\left(  \mathbf{h},P^{\ast}\right)  \equiv
\bigcup\limits_{\substack{\pi\in\Pi\\\sum_{m=1}^{M}\left\|  \mathbf{p}%
_{m}\right\|  _{1}\leq P^{\ast}}}\left\{  \mathbf{R}:R_{\pi\left(  m\right)
}\leq\tilde{R}_{\pi\left(  m\right)  }\;,m\in\mathcal{M}\right\}
\]
Here, $\mathbf{R}=[R_{1},...,R_{M}]^{T}$ and $\tilde{R}_{\pi\left(  m\right)
}$ is the rate of user $\pi\left(  m\right)  $ defined in (\ref{rate_bc}).

%
%
%
%
%
%
%
%
%
%
%
%
%
%
%
%
%
%
%
%
%
%
%
%
%
%
%
%
%

\section{OFDM single user DLC}

\label{sec:su}

\subsection{An implicit formulation of the single user OFDM DLC}

First let us define the single user DLC $C_{d}$ for OFDM, which is a special
case of parallel fading channels. The user index $m$ is omitted in this section.

\begin{definition}
\label{def:dlc_su} A rate $R^{\ast}$ is achievable with limited delay (zero
outage) under a long term power constraint $P^{\ast}$ if and only if for any
fading state $\mathbf{h}$ there exists a power allocation $\mathbf{p}^{\prime
}$ solving
\begin{equation}
\min||\mathbf{p}^{\prime}||_{1}\text{\quad subj. to\quad}R^{\ast}\leq
R(\mathbf{h},\mathbf{p}^{\prime}) \label{eq:dlc_su}%
\end{equation}
and
\[
\mathbb{E}_{\mathbf{h}}\left(  ||\mathbf{p}^{\prime}||_{1}\right)  \leq
P^{\ast}.
\]
Furthermore $R^{\ast}$ is called the \emph{delay limited capacity}
$C_{d}(P^{\ast})$ if and only if
\[
\mathbb{E}_{\mathbf{h}}\left(  ||\mathbf{p}^{\prime}||_{1}\right)  =P^{\ast}.
\]
\end{definition}

Obviously, the delay-limited capacity is smaller than the ergodic capacity
under the same average power constraint because the requested rate has to be
supported in every fading state. Observe that we choose to state the DLC
region as a definition rather than a theorem as done in \cite{hanly_98_inf}.
There it is shown that our definition coincides with rates that guarantee
arbitrary small erroneous decoding probability (dependent on the coding delay)
for any jointly stationary, ergodic fading process when the codewords can be
chosen as a function of the realization of the fading process. This can be
interpreted as a performance limit for traffic where the information has to be
delivered within one coding frame for almost all fading realizations and the
channel varies slowly enough such that the transmitter can track the channel.
Moreover, by our definition we see that since the sum power minimization
problem has to be solved in every fading step the DLC represents also a
performance limit for practical minimum sum power algorithms. We will present
an example in Sec. \ref{sec:ofdma}.

The single user DLC was already examined by several authors in the context of
systems with multiple antennas and parallel independent fading channels
\cite{biglieri_01_inf,jorswieck_05_itg}. Nevertheless, we re-derive the DLC
here using the principle of \emph{rate water-filling}, which leads to an
interesting perspective. This characterization of the DLC will be used later
on to derive lower bounds on the OFDM broadcast channel DLC region in Section
\ref{sec:ofdma}. Denoting the rate on subcarrier $k$ as $r_{k}$, the
optimization problem for each fading state in (\ref{eq:dlc_su}) is equivalent
to
\begin{equation}
\min\sum_{k=1}^{K}p_{k}-\lambda\sum_{k=1}^{K}r_{k}\quad\text{subj. to}%
\quad\sum_{k=1}^{K}r_{k}\geq KC_{d}.\nonumber
\end{equation}
Using the relation between power and rate on each subcarrier this can be
expressed as
\begin{equation}
\min\sum_{k=1}^{K}\left(  \frac{e^{r_{k}}-1}{h_{k}}-\lambda r_{k}\right)
,\nonumber
\end{equation}
where $\lambda>0$ is to be chosen so that $\sum_{k=1}^{K}r_{k}=KC_{d}$. The
resulting optimality conditions are given by
\begin{equation}
\left[  \frac{e^{r_{k}}}{h_{k}}-\lambda\right]  ^{-}=0\quad\forall k,\quad
\sum_{k=1}^{K}r_{k}=KC_{d} \label{eqn:kkt_dlc1}%
\end{equation}
with $\left[  \cdot\right]  ^{-}=\min\left\{  \cdot,0\right\}  $. Note that by
taking logarithms and solving for $r_{k}$ eqn. (\ref{eqn:kkt_dlc1}) can be
interpreted as a water-filling solution with respect to the rates $r_{k}$.
Combining the optimality conditions for all $K$ subcarriers and solving for
the Lagrangian multiplier $\lambda$ yields
\begin{equation}
\lambda=\frac{\exp\left(  \frac{C_{d}K}{d_{\mathbf{h}}}\right)  }{\prod
_{k\in\mathcal{D}\left(  \mathbf{h}\right)  }h_{k}^{d_{\mathbf{h}}^{-1}}},
\label{eqn:lambda}%
\end{equation}
where the random variable $\mathcal{D}\left(  \mathbf{h}\right)
\subseteq\mathcal{K}$ denotes the set of \emph{active} subcarriers and
$d_{\mathbf{h}}:=\left|  \mathcal{D}\left(  \mathbf{h}\right)  \right|  $.
Note that the allocated power is given by $p_{k}=\lambda-h_{k}^{-1}$ for any
$k\in\mathcal{D}\left(  \mathbf{h}\right)  $ and zero otherwise. Substituting
(\ref{eqn:lambda}) in the average power expression given by
\begin{equation}
P^{\ast}=\mathbb{E}_{\mathbf{h}}\left(  \sum_{k=1}^{K}p_{k}(\lambda)\right)
=\mathbb{E}_{\mathbf{h}}\left(  \sum_{k=1}^{K}\left[  \lambda-\frac{1}{h_{k}%
}\right]  ^{+}\right)
\end{equation}
with $\left[  \cdot\right]  ^{+}=\max\left\{  \cdot,0\right\}  $ we obtain the
single user OFDM delay limited capacity $C_{d}$ with power constraint
$P^{\ast}$:
\begin{equation}
P^{\ast}=\mathbb{E}_{\mathbf{h}}\left(  \frac{d_{\mathbf{h}}\exp\left(
\frac{C_{d}K}{d_{\mathbf{h}}}\right)  }{K\prod_{k\in\mathcal{D}\left(
\mathbf{h}\right)  }h_{k}^{d_{\mathbf{h}}^{-1}}}\right)  -\frac{1}%
{K}\mathbb{E}_{\mathbf{h}}\left(  \sum_{k\in\mathcal{D}\left(  \mathbf{h}%
\right)  }\frac{1}{h_{k}}\right)  \label{eqn:dlc1}%
\end{equation}
Since the denominator in (\ref{eqn:dlc1}) can be bounded by a constant, the
delay limited capacity $C_{d}$ is greater than zero if and only if
\begin{equation}
\int\limits_{\mathbb{R}_{+}^{K}}\frac{1}{\prod_{k\in\mathcal{D}\left(
\mathbf{h}\right)  }h_{k}^{d_{\mathbf{h}}^{-1}}}\,dF_{\mathbf{h}}\left(
\mathbf{h}\right)  <\infty. \label{eqn:regular}%
\end{equation}
Here, $F_{\mathbf{h}}$ denotes the joint fading distribution function. The
class of fading distributions for which (\ref{eqn:regular}) holds is called
$\emph{regular}$ in \cite{biglieri_01_inf}. It will become apparent in the
following that the correlation structure of the channel gains in OFDM provides
the main challenge in proving and analyzing regularity according to
(\ref{eqn:regular}).

\subsection{Suboptimal power allocation strategies}

Let us introduce a suboptimal power allocation for the single-user case that
is used in \ref{sec:ofdma}. It is evident from the expression for the single
user DLC that the major difficulty is the rate water-filling operation for
each channel realization. To circumvent this difficulty which results in a
prohibitive complexity for the multi-user case we introduce \emph{the notion
of rate water-filling} for average channel realizations. The idea is to use
simply the information which subcarrier is the best, the second best and so on
and to allocate fixed rate budgets to the in that way ordered subcarriers. For
the analysis we need the following definitions. For a given vector
$\mathbf{h}$ of real elements let us introduce the total ordering
\[
h_{k\left[  K\right]  }\geq h_{k\left[  K-1\right]  }\geq\ldots\geq
h_{k\left[  1\right]  },
\]
i.e. $h_{k\left[  1\right]  }$ is the minimum value and $h_{k\left[  M\right]
}$ is the maximum value. If $\mathbf{h}$ is a random variable then the
distribution of $h_{k\left[  p\right]  }$ is known to be the $p$\emph{-th
order statistics}. The the $p$-th order statistics can be explicitely given
for $K$ independent random variables $h_{k}$ with distribution $F$ and density
$f$ from standard books. The $p$-th order density is given by%
\begin{equation}
f_{h_{k\left[  p\right]  }}\left(  x\right)  =Kf\left(  x\right)  \binom
{K-1}{p}F^{p-1}\left(  x\right)  \left(  1-F\left(  x\right)  \right)  ^{K-p}
\label{eqn:order_density}%
\end{equation}
Based on the order information and the distribution we can now deduce a fixed
rate allocation on the subcarriers, avoiding the water-filling procedure in
each fading state. The idea is to allocate a fixed rate budget to the $p$-th
ordered subcarrier. Defining the terms
\[
\zeta_{p}=\int_{0}^{\infty}h^{-1}\,dF_{h_{k\left[  p\right]  }}\left(
h\right)
\]
and using these factors the optimal rate allocation is now given by solving
the optimization problem%
\[
\min_{R_{k\left[  p\right]  }\geq0}\,\frac{e^{R_{k\left[  p\right]  }}%
-1}{\zeta_{p}^{-1}}-\lambda R_{k},\quad\sum_{k=1}^{K}R_{k\left[  p\right]
}=KC_{d},
\]
which we have already solved by rate water-filling given by
\[
\left[  \frac{\log\left(  e\right)  e^{R_{k\left[  p\right]  }}}{\zeta
_{p}^{-1}}-\lambda\right]  ^{-}=0,\quad\sum_{p=1}^{K}R_{k\left[  p\right]
}=KC_{d}.
\]
Obviously, this suboptimal scheme has some impact on the \emph{feedback
protocol}: Since only the order statistics are exploited, it seems to suffice
to feed back the ordering of the subcarriers. This affords much less feedback
capacity than perfect channel knowledge would require. However, it is not
straightforward to translate rate water-filling to a power allocation
strategy, since for achieving a certain rate the channel has to be perfectly
known. Allocating fixed power budgets is possible. The consequence is, that
since the $p$-th order channel is a random variable, mutual information
becomes a random variable once again not guaranteeing a certain rate in each
state. However, the variance becomes much smaller.

\begin{figure}[ptb]
\begin{center}
\includegraphics[width=10cm]{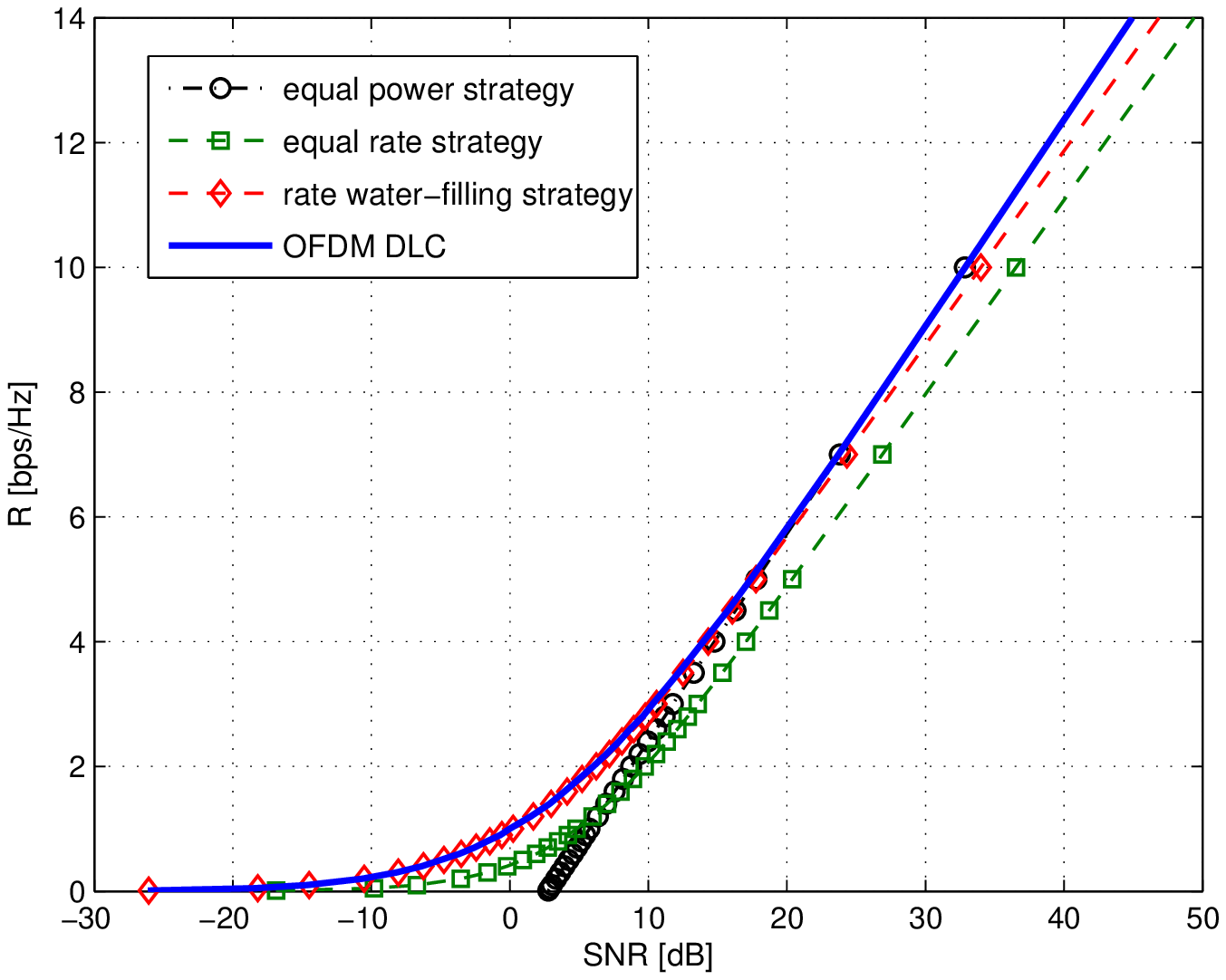}
\end{center}
\caption{OFDM DLC and lower bounds for L=K=16.}%
\label{fig:dlc1_ord}%
\end{figure}

There is an interesting second power allocation where the powers asserted to
the subcarriers are all the same. It is easy to see that the allocation
according to
\begin{equation}
p_{k}=\frac{e^{R}}{\mathbb{E}_{\mathbf{h}}\left(  \prod_{k=1}^{K}h_{k}%
^{-1/K}\right)  },\quad k\in\mathcal{K}, \label{eqn:pclaw}%
\end{equation}
always leads to a rate higher than the requested rate with equality in the
high SNR region. Hence, this is also a suboptimal solution. The bounds are
illustrated in Fig. \ref{fig:dlc1_ord}.

It is of great interest to understand the impact of the delay spread $L$ and
the power delay profile $\boldsymbol{\sigma}$ as well as the fading
distribution on the OFDM delay limited capacity. In case of the OFDM broadcast
channel, an analytical characterization is nearly impossible. Thus we carry
out an analysis for the single user OFDM DLC in the following. Note that this
matches the behavior on the axes of the OFDM BC DLC region where only one user
is active. Hence the results give insights for the broadcast case as well.
Since the expression in (\ref{eqn:dlc1}) is still very complicated, we focus
on the behavior in the low and the high SNR regime and carry out a detailed analysis.

\section{Impact of system parameters}

\label{sec:impact}

\subsection{Scaling in low SNR}

\subsubsection{Impact of delay spread and fading distribution}

First, we characterize the first and second order behavior at low SNR. For
ease of notation we define $h_{\infty}:=\left\|  \mathbf{h}\right\|  _{\infty
}$.

\begin{proposition}
\label{prop:low2} Suppose that $\mathbb{E}_{\mathbf{h}}\left(  h_{\infty}%
^{-1}\right)  <\infty$. Then, the following limit holds:
\begin{equation}
\lim_{P^{\ast}\rightarrow0}\frac{C_{d}\left(  P^{\ast}\right)  }{P^{\ast}%
}=\frac{1}{\mathbb{E}_{\mathbf{h}}\left(  h_{\infty}^{-1}\right)  }
\label{eq:first_order_low_snr}%
\end{equation}
\end{proposition}

\begin{proof}
Our starting point is (\ref{eqn:dlc1}) where we use the McLaurin-expansion of
the exponential function $\exp\left(  x\right)  =1+x+o\left(  x\right)  $ up
to the linear term to obtain a lower bound on the required power $P^{\ast}$.

Fix now $\epsilon>0$ and use the following strategy: set the 'virtual' channel
gain of any subcarrier $k$ with $h_{k}\geq h_{\infty}-\epsilon$ to$\ h_{\infty
}$. Denote by $\chi_{\mathbf{h}}\left(  \epsilon\right)  $ the multiplicity of
the number of subchannels that are assigned the maximum channel gain by this
strategy for any channel realization. Using (\ref{eqn:dlc1}) we have for
sufficiently small $P^{\ast}$ an upper bound on $C_{d}$ which is given by
\begin{align}
P^{\ast}  &  \geq\mathbb{E}_{\mathbf{h}}\left(  \frac{\chi_{\mathbf{h}}\left(
\epsilon\right)  +C_{d}K}{Kh_{\infty}}\right)  -\frac{1}{K}\mathbb{E}%
_{\mathbf{h}}\left(  \frac{\chi_{\mathbf{h}}\left(  \epsilon\right)
}{h_{\infty}}\right) \\
&  =C_{d}\mathbb{E}_{\mathbf{h}}\left(  \frac{1}{h_{\infty}}\right)
\end{align}
and we obtain
\begin{equation}
C_{d}\leq\frac{P^{\ast}}{\mathbb{E}_{\mathbf{h}}\left(  h_{\infty}%
^{-1}\right)  }%
\end{equation}
for any $\epsilon>0$.

For the upper bound on $P^{\ast}$ fix $\epsilon^{\prime}>0$. This time set the
number of supported subcarriers to one and support one of the set with maximum
channel gain. Then we have for sufficiently small $P^{\ast}$
\begin{equation}
P^{\ast}\leq\mathbb{E}_{\mathbf{h}}\left(  \frac{1+\left(  1+\epsilon^{\prime
}\right)  C_{d}K}{Kh_{\infty}}\right)  -\frac{1}{K}\mathbb{E}_{\mathbf{h}%
}\left(  \frac{1}{h_{\infty}}\right)
\end{equation}
Hence we have
\begin{equation}
C_{d}\geq\frac{P^{\ast}}{\left(  1+\epsilon^{\prime}\right)  \mathbb{E}%
_{\mathbf{h} }\left(  h_{\infty}^{-1}\right)  }%
\end{equation}
for any $\epsilon^{\prime}>0$. Combining both the lower and upper bound yields
the desired result.
\end{proof}

This quantity also reveals the minimum energy per bit, at which reliable
communication is possible under a limited delay. This is in analogy to
\cite{Ver:1990}, where this quantity was derived for the ergodic capacity of a
Gaussian channel. Moreover, the lemma states that albeit it is generally
suboptimal, serving one of the best subcarriers becomes optimal in the low SNR
region. This can be easily seen since the multiplicity $\chi_{\mathbf{h}%
}(\epsilon)$ of the attained maximum channel gain vanishes in the expressions
for the lower and upper bound.

To gain more insights, the remaining sub-linear term defined by
\begin{equation}
\Delta_{d}\left(  P^{\ast}\right)  :=C_{d}^{\prime}\left(  0\right)  P^{\ast
}-C_{d}\left(  P^{\ast}\right)
\end{equation}
is calculated next. The following proposition tells us that while for the
linear term it did not matter if the distribution contains point masses it
does matter for the sub-linear term:

\begin{proposition}
\label{prop:low3}The following limit for the sub-linear term holds:
\begin{equation}
\underset{P^{\ast}\rightarrow0}{\lim}\frac{\Delta_{d}\left(  P^{\ast}\right)
}{\left(  P^{\ast}\right)  ^{2}}=\frac{K\mathbb{E}_{\mathbf{h}}\left(
\chi_{\mathbf{h}}^{-1}h_{\infty}^{-1}\right)  }{2\mathbb{E}_{\mathbf{h}}%
^{3}\left(  h_{\infty}^{-1}\right)  } \label{eq:second_order_low_snr}%
\end{equation}
Here, $\chi_{\mathbf{h}}$ is the (random) multiplicity of subchannels with
maximum channel gain.
\end{proposition}

\begin{proof}
Our starting point is again (\ref{eqn:dlc1}) where we now use the
McLaurin-expansion of the exponential function up to the quadratic term, i.e.
$\exp\left(  x\right)  =1+x+0.5x^{2}+o\left(  x^{2}\right)  $.

By the same strategy as above fix $\epsilon>0$ and set the ''virtual'' channel
gain of any subcarrier $k$ with $h_{k}\geq h_{\infty}-\epsilon$ to$\ h_{\infty
}$. Then, we obtain for sufficiently small $P^{\ast}$%
\begin{equation}
P^{\ast}\geq\mathbb{E}_{\mathbf{h}}\left(  \frac{C_{d}+0.5C_{d}^{2}%
K\chi_{\mathbf{h}}^{-1}\left(  \epsilon\right)  }{h_{\infty}}\right)  .
\end{equation}
The equation is an upward open parabola in $C_{d}$ where one zero is negative
and one is positive where the latter is increasing in $P^{\ast}$. Solving this
equation for $C_{d}$ yields the inequality
\begin{equation}%
\begin{split}
C_{d}\left(  P^{\ast}\right)   &  \leq-\frac{\mathbb{E}_{\mathbf{h}}\left(
h_{\infty}^{-1}\right)  }{K\mathbb{E}_{\mathbf{h}}\left(  \chi_{\mathbf{h}%
}^{-1}\left(  \epsilon\right)  h_{\infty}^{-1}\right)  }\\
&  +\sqrt{\frac{\mathbb{E}_{\mathbf{h}}^{2}\left(  h_{\infty}^{-1}\right)
}{K^{2}\mathbb{E}_{\mathbf{h}}^{2}\left(  \chi_{\mathbf{h}}^{-1}\left(
\epsilon\right)  h_{\infty}^{-1}\right)  }+\frac{2P^{\ast}}{K\mathbb{E}%
_{\mathbf{h}}\left(  \chi_{\mathbf{h}}^{-1}\left(  \epsilon\right)  h_{\infty
}^{-1}\right)  }}.
\end{split}
\end{equation}
Expanding the square root function yields
\begin{equation}
C_{d}\left(  P^{\ast}\right)  \leq\frac{1}{\mathbb{E}_{\mathbf{h}}\left(
h_{\infty}^{-1}\right)  }P^{\ast}-\frac{K\mathbb{E}_{\mathbf{h}}\left(
\chi_{\mathbf{h}}^{-1}\left(  \epsilon\right)  h_{\infty}^{-1}\right)
}{2\left(  1+\epsilon\right)  \mathbb{E}_{\mathbf{h}}^{3}\left(  h_{\infty
}^{-1}\right)  }\left(  P^{\ast}\right)  ^{2}. \label{eq:root_expansion}%
\end{equation}
Subtracting the first order expression (\ref{eq:first_order_low_snr}) from
(\ref{eq:root_expansion}) we arrive for some $\epsilon^{\prime}>0$ at
\begin{align}
\Delta_{d}\left(  P^{\ast}\right)   &  \geq\frac{K\mathbb{E}_{\mathbf{h}%
}\left(  \chi_{\mathbf{h}}^{-1}\left(  \epsilon\right)  h_{\infty}%
^{-1}\right)  }{2\left(  1+\epsilon\right)  \mathbb{E}_{\mathbf{h}}^{3}\left(
h_{\infty}^{-1}\right)  }\left(  P^{\ast}\right)  ^{2}\nonumber\\
&  \geq\frac{K\mathbb{E}_{\mathbf{h}}\left(  \chi_{\mathbf{h}}^{-1}h_{\infty
}^{-1}\right)  -\epsilon^{\prime}}{2\left(  1+\epsilon\right)  \mathbb{E}%
_{\mathbf{h}}^{3}\left(  h_{\infty}^{-1}\right)  }\left(  P^{\ast}\right)
^{2} \label{eq:upper_bound_low_snr_second_order}%
\end{align}
and thus have established a lower bound on $\Delta_{d}(P^{\ast})$ for any
$\epsilon,\epsilon^{\prime}>0$. The last inequality
(\ref{eq:upper_bound_low_snr_second_order}) follows from the following
argument (which is frequently used in the sequel): observe that $\chi
_{\mathbf{h}}\left(  \epsilon\right)  \geq1$ and, almost surely with respect
to the fading distribution, for any realization $\mathbf{h}$ we have
\begin{equation}
\chi_{\mathbf{h}}^{-1}\left(  \epsilon\right)  h_{\infty}^{-1}\rightarrow
\chi_{\mathbf{h}}^{-1}h_{\infty}^{-1},\quad\epsilon\rightarrow0,
\end{equation}
and provided that $\mathbb{E}_{\mathbf{h}}\left(  \chi_{\mathbf{h}}%
^{-1}h_{\infty}^{-1}\right)  \leq\mathbb{E}_{\mathbf{h}}\left(  h_{\infty
}^{-1}\right)  <\infty$ we obtain by dominated convergence
\cite{kolmogorov_70}:
\begin{equation}
\lim_{\epsilon\rightarrow0}\mathbb{E}_{\mathbf{h}}\left(  \chi_{\mathbf{h}%
}^{-1}\left(  \epsilon\right)  h_{\infty}^{-1}\right)  =\mathbb{E}%
_{\mathbf{h}}\left(  \chi_{\mathbf{h}}^{-1}h_{\infty}^{-1}\right)
\end{equation}
In analogy to the derivation of the first order behavior we get
\begin{equation}
\Delta_{d}\left(  P^{\ast}\right)  \leq\frac{\left(  1+\epsilon^{\prime
}\right)  K\mathbb{E}_{\mathbf{h}}\left(  \chi_{\mathbf{h}}^{-1}h_{\infty
}^{-1}\right)  }{2\mathbb{E}_{\mathbf{h}}^{3}\left(  h_{\infty}^{-1}\right)
}\left(  P^{\ast}\right)  ^{2}. \label{eq:lower_bound_low_snr_second_order}%
\end{equation}
Combining (\ref{eq:upper_bound_low_snr_second_order}) and
(\ref{eq:lower_bound_low_snr_second_order}) leads to the desired result.
\end{proof}

\begin{corollary}
\label{cor:1}Since $\chi_{\mathbf{h}}(\epsilon)\geq1$ a simple upper bound on
the second order term in (\ref{eq:second_order_low_snr}) is given by
\begin{equation}
\underset{P^{\ast}\rightarrow0}{\lim\sup}\frac{\Delta_{d}\left(  P^{\ast
}\right)  }{\left(  P^{\ast}\right)  ^{2}}\leq\frac{K}{2\left[  \mathbb{E}%
_{\mathbf{h}}\left(  h_{\infty}^{-1}\right)  \right]  ^{2}}.
\end{equation}
\end{corollary}

Note that the bound from Corollary \ref{cor:1} is consistent with the result
in \cite{wunder_06_itw} where it is shown that
\begin{equation}
C_{d}\left(  P^{\ast}\right)  \sim\frac{1}{K}\log\left(  1+\frac{KP^{\ast}%
}{\mathbb{E}_{\mathbf{h}}\left(  h_{\infty}^{-1}\right)  }\right)  ,\quad
P^{\ast}\rightarrow0. \label{eqn:logf}%
\end{equation}
This can be easily checked by differentiating the expression in
(\ref{eqn:logf}) twice.

\begin{proposition}
\label{prop:low4} Suppose that the joint fading distribution is absolute
continuous. Then, the following limit holds:%
\begin{equation}
\underset{P^{\ast}\rightarrow0}{\lim}\frac{\Delta_{d}\left(  P^{\ast}\right)
}{\left(  P^{\ast}\right)  ^{2}}=\frac{K}{\left[  \mathbb{E}_{\mathbf{h}
}\left(  h_{\infty}^{-1}\right)  \right]  ^{2}}%
\end{equation}
\end{proposition}

\begin{proof}
We have to prove that the set of events where the maximum is taken on by more
than one subcarrier has probability zero.

By the absolute continuity of the joint fading distribution (which is
preserved under unitary mappings) this is equivalent to show that the set that
contains all events where the maximum is not unique has Lebesgue measure zero.
To see this we consider sets of the form
\[
\left\{  \mathbf{h}\in\mathbb{R}_{+}^{K}:\bigcup\limits_{1\leq i\leq
K}\left\{  h_{\pi\left(  1\right)  }=\ldots=h_{\pi\left(  i\right)  }%
,h_{\pi\left(  i+1\right)  },\ldots,h_{\pi\left(  K\right)  }\right\}
\right\}
\]
where $\pi$ is the permutation that yields an decreasing order among the
channel gains
\[
h_{\pi\left(  1\right)  }\geq h_{\pi\left(  2\right)  }\geq\ldots\geq
h_{\pi\left(  K\right)  }%
\]
Observe that any set has measure zero. Since there can only be $2^{K}$ such
possible sets the union of these sets has measure zero as well. Now, as our
regarded set is in the union of the constructed set, the set has also measure zero.
\end{proof}

Hence, appealing to Prop. \ref{prop:low2}, \ref{prop:low3} and \ref{prop:low4}
the forthcoming analysis reduces to the study of the expected maximum of the
channel gains. However, the expressions do not show how the DLC depends on the
system parameters which we investigate by means of an asymptotic analysis, i.e.
for large $L,K$. This analysis turns out to be quite accurate even for very
small $L$.

\begin{remark}
It is important to note that for the asymptotic analysis we will let go $L$
and $K$, $K\geq L$, to infinity which is indicated by the index $n$, i.e.
$K_{n},L_{n}\rightarrow\infty$ as $n\rightarrow\infty$. We assume that for all
$L_{n}$ the complex path gain vectors $\mathbf{\tilde{c}}_{n}=[\hat{c}%
_{1},\ldots,\hat{c}_{L_{n}}]^{T}$ are defined on the same probability space,
i.e. to each $\mathbf{\tilde{c}}_{n}\in\mathbb{C}^{L_{n}}$ there is by means
of (\ref{eqn:channel}) $\mathbf{\tilde{h}}_{n}=[\hat{h}_{1},\ldots,\hat{h}%
_{K}]^{T}\in\mathbb{C}^{K_{n}}$. While the distribution does not change for
the $L_{n-1}$ first random variables of the vector $\mathbf{\tilde{c}}_{n}$
the distribution of the $K_{n-1}$ first random variables in $\mathbf{\tilde
{h}}_{n}$ might change since they depend on the FFT structure. We have to keep
this in mind for the forthcoming analysis.
\end{remark}

It was shown in \cite{wunder_06_inf} that the (continuous) maximum value of
the frequency response equals $\log\left(  L\right)  $ with large probability
even for moderate $L$ for the following distributions:
\begin{align}
&  \tilde{c}_{i}\text{ independent and iid (both in real and imaginary parts)
}\forall i,\nonumber\\
&  \mathbb{E}_{\mathbf{h}}\left(  \Re e^{2}\left(  \tilde{c}_{1}\right)
\right)  =.5/L,\mathbb{E}_{\mathbf{h}}\left(  e^{j\omega\Re e\left(  \tilde
{c}_{1}\right)  }\right)  =e^{-.25L^{-1}\omega^{2}+\sum_{l=3}^{5}a_{l}%
\omega^{l}+O\left(  \omega^{6}\right)  },\label{eqn:dist1}\\
&  \text{for all }\left|  \omega\right|  \leq d,\quad d>0,a_{3},a_{4},a_{5}%
\in\mathbb{C}\nonumber
\end{align}
Here, $\Re e\left(  c\right)  $ denotes the real part of the complex number
$c$ ($\Im m$ is the imaginary part). Note that the condition on the
characteristic function of the real part of the path gains implies finiteness
of the moments up to order six of the corresponding distribution. The
following theorem proves that the $\log\left(  L\right)  $ result holds also
for the maximum of the sampled frequency response regardless of the sampling
set. Furthermore, it reveals the exceptional role of uniform PDP.

\begin{theorem}
\label{theorem:dist}Suppose that the fading distribution belongs to
$\mathcal{F}_{L}$ then we have
\[
\Pr\left(  \log\left(  L\right)  -g\left(  L\right)  \leq h_{\infty}\leq
\log\left(  L\right)  +g\left(  L\right)  \right)  =1-O\left(  \log
^{-4}\left(  L\right)  \right)
\]
with $g\left(  L\right)  =4\log\left[  \log\left(  L\right)  \right]  $ for
large $L$ and arbitrary $K\geq L$. Furthermore, the upper bound
\[
\Pr\left(  h_{\infty}\geq\log\left(  L\right)  +g\left(  L\right)  \right)
\leq O\left(  \log^{-4}\left(  L\right)  \right)
\]
also holds when the PDP is non-uniform.
\end{theorem}

\begin{proof}
see Appendix \ref{proof_of_theorem:dist}.
\end{proof}

We can apply this result to the DLC assuming uniform PDP where we have to show
that from the convergence in probability given in Theorem \ref{theorem:dist}
follows convergence of the expected maximum of the channel gains. This can be
achieved if the set of distributions is somewhat more restricted compared to
(\ref{eqn:dist1}) in the sense that the their behavior in the neighborhood of
the zero is ''sufficiently well''. By this we mean, that the distribution
function of $c_{1}$ is Lipschitz continuous in an $\epsilon$-neighborhood of
the zero. Let us denote this class of fading distributions by:
\begin{align*}
\mathcal{F}_{\mathrm{Lo}}\left(  k_{s}\right)   &  :=\left\{  \text{the
conditions (\ref{eqn:dist1}) on }\tilde{c}_{i}\text{ hold }\forall i\right. \\
&  \left.  \mathbf{h}\in\mathbb{R}_{+}^{K}\text{ is generated from
}\mathbf{\tilde{c}}\in\mathbb{C}_{+}^{L}\text{ by means of (\ref{eqn:channel}%
)}\right. \\
&  \left.  F_{c_{1}}\left(  x\right)  \leq k_{s}x,0\leq x<\epsilon
,k_{s},\epsilon\in\mathbb{R}_{++}\right\}
\end{align*}
The Lipschitz continuity is essential in the next Lemma.

\begin{lemma}
\label{lemma:low_snr}Suppose that the fading distribution belongs to
$\mathcal{F}_{\mathrm{Lo}}\left(  k_{s}\right)  $ then the following limit
holds for sufficiently large $L$ and arbitrary $K\geq L$:
\[
\underset{P^{\ast}\rightarrow0}{\lim}\frac{C_{d}\left(  P^{\ast}\right)
}{P^{\ast}}=\log\left(  L\right)  +O\left(  \log\left[  \log\left(  L\right)
\right]  \right)
\]
Furthermore, the DLC is maximized (with respect to the leading order term) by
uniform PDP (''order-optimal'').
\end{lemma}

\begin{proof}
The last statement follows from Theorem \ref{theorem:dist} and, hence, we
assume uniform PDP. By the same theorem there exist constants $\gamma
,\kappa\in\mathbb{R}_{++}$ so that
\begin{align*}
&  \Pr\left(  \log\left(  L\right)  -\gamma\log\left[  \log\left(  L\right)
\right]  \leq h_{\infty}\leq\log\left(  L\right)  +\gamma\log\left[
\log\left(  L\right)  \right]  \right) \\
&  \geq1-\frac{\kappa}{\log^{\gamma}\left(  L\right)  }%
\end{align*}
for sufficiently large $L$. Setting $\epsilon_{-}=1-\epsilon$ and
$\epsilon_{+}=1+\epsilon$ where $\epsilon:=\gamma\log\left[  \log\left(
L\right)  \right]  /\log\left(  L\right)  $ the expectation can be written as
\begin{align*}
&  \mathbb{E}_{\mathbf{h}}\left(  h_{\infty}^{-1}\right) \\
&  =\mathbb{E}_{\mathbf{h}}\left(  \left.  h_{\infty}^{-1}\right|  h_{\infty
}\in\left[  \epsilon_{-}\log\left(  L\right)  ,\epsilon_{+}\log\left(
L\right)  \right]  \right)  \,\Pr\left(  h_{\infty}\in\left[  \epsilon_{-}%
\log\left(  L\right)  ,\epsilon_{+}\log\left(  L\right)  \right]  \right) \\
&  +\mathbb{E}_{\mathbf{h}}\left(  \left.  h_{\infty}^{-1}\right|  h_{\infty
}<\epsilon_{-}\log\left(  L\right)  \right)  \,\Pr\left(  h_{\infty}%
<\epsilon_{-}\log\left(  L\right)  \right) \\
&  +\mathbb{E}_{\mathbf{h}}\left(  \left.  h_{\infty}^{-1}\right|  h_{\infty
}>\epsilon_{+}\log\left(  L\right)  \right)  \,\Pr\left(  h_{\infty}%
>\epsilon_{+}\log\left(  L\right)  \right)  .
\end{align*}
Note that the crucial part is to derive an upper bound on the conditional
expectation in the second term on the RHS of the last equation, i.e. when the
maximum $h_{\infty}$ is small. We will show now that the conditional
expectation is bounded too. Proceeding with the standard inequality
$\mathbb{E}\left(  \left|  X\right|  \right)  \leq1+\sum_{n=1}^{\infty}%
\Pr\left(  \left|  X\right|  \geq n\right)  $ and using the ''trick'' that we
can apply it to the conditional probability measure as well, the second term
is given by
\[
\mathbb{E}_{\mathbf{h}}\left(  \left.  h_{\infty}^{-1}\right|  h_{\infty
}<\epsilon_{-}\log\left(  L\right)  \right)  \leq1+\sum_{i=1}^{+\infty}%
\frac{\Pr\left(  \left\{  h_{\infty}\leq\frac{1}{i}\right\}  \cap\left\{
h_{\infty}<\epsilon_{-}\log\left(  L\right)  \right\}  \right)  }{\Pr\left(
\left\{  h_{\infty}<\epsilon_{-}\log\left(  L\right)  \right\}  \right)  }.
\]
Using the inequality $\left\|  \mathbf{c}\right\|  _{1}=\left\|
\mathbf{h}\right\|  _{1}/K\leq$ $h_{\infty}$, and $\left\|  \mathbf{c}%
\right\|  _{\infty}\leq\left\|  \mathbf{c}\right\|  _{1}$, define $P\left(
x\right)  :=\Pr\left(  \left\{  h_{\infty}<x\right\}  \right)  $ and we obtain
by independence of the path gains
\begin{align}
&  \mathbb{E}_{\mathbf{h}}\left(  \left.  h_{\infty}^{-1}\right|  h_{\infty
}<\epsilon_{-}\log\left(  L\right)  \right) \nonumber\\
&  \leq1+\sum_{i=1}^{1+\left\lfloor k_{s}\right\rfloor }\frac{P\left(
\frac{1}{i}\right)  }{P\left(  \epsilon_{-}\log\left(  L\right)  \right)
}+\frac{1}{P\left(  \epsilon_{-}\log\left(  L\right)  \right)  }%
\sum_{i=2+\left\lfloor k_{s}\right\rfloor }^{+\infty}\Pr\left(  \bigcap
_{l=1}^{L}\left\{  c_{l}\leq\frac{1}{i}\right\}  \right)  \label{eqn:ks}%
\end{align}
where $\left\lfloor k_{s}\right\rfloor $ is the greatest natural number below
$k_{s}$. Hence
\begin{align}
\text{RHS of (\ref{eqn:ks})}  &  =2+\left\lfloor k_{s}\right\rfloor +\frac
{1}{P\left(  \epsilon_{-}\log\left(  L\right)  \right)  }\sum
_{i=2+\left\lfloor k_{s}\right\rfloor }^{+\infty}F_{c_{1}}^{L}\left(  \frac
{1}{i}\right) \nonumber\\
&  \leq2+\left\lfloor k_{s}\right\rfloor +\frac{k_{s}^{L}}{P\left(
\epsilon_{-}\log\left(  L\right)  \right)  }\sum_{i=2+\left\lfloor
k_{s}\right\rfloor }^{+\infty}\frac{1}{i^{L}}. \label{eqn:ks1}%
\end{align}
In the second step we assumed Lipschitz continuity of the path gain
distribution function $F_{c_{1}}$ with Lipschitz constant below or equal
$k_{s}$, i.e. $F_{c_{1}}\left(  x\right)  \leq k_{s}x$. The series in
(\ref{eqn:ks1}) can be upper bounded as follows:
\begin{align*}
\sum_{i=2+\left\lfloor k_{s}\right\rfloor }^{+\infty}\frac{1}{i^{L}}  &
\leq\sum_{i=2+\left\lfloor k_{s}\right\rfloor }^{+\infty}\int_{i-1}^{i}%
\frac{1}{x^{L}}\,dx\\
&  =\int_{1+\left\lfloor k_{s}\right\rfloor }^{+\infty}\frac{1}{x^{L}}\,dx\\
&  =\frac{1}{-L+1}\left[  x^{-L+1}\right]  _{1+\left\lfloor k_{s}\right\rfloor
}^{\infty}\\
&  \leq\frac{1}{L-1}k_{s}^{-L+1}%
\end{align*}

Hence, we obtain finally
\[
\text{RHS of (\ref{eqn:ks1})}\leq2+\left\lfloor k_{s}\right\rfloor +\frac
{1}{P\left(  \epsilon_{-}\log\left(  L\right)  \right)  }\frac{k_{s}}{L-1}.
\]

The last term is finite for $L>1$. Hence, we have the upper bound%
\begin{align*}
\mathbb{E}_{\mathbf{h}}\left(  h_{\infty}^{-1}\right)   &  \leq\epsilon
_{-}^{-1}\log^{-1}\left(  L\right)  +\frac{\left(  2+\left\lfloor
k_{s}\right\rfloor \right)  \kappa}{\log^{\gamma}\left(  L\right)  }%
+\frac{k_{s}}{L-1}+\epsilon_{+}^{-1}\log^{-1}\left(  L\right)  \frac{\kappa
}{\log^{\gamma}\left(  L\right)  }\\
&  =\epsilon_{-}^{-1}\log^{-1}\left(  L\right)  \,\left(  1+\frac{\left(
2+\left\lfloor k_{s}\right\rfloor \right)  \epsilon_{-}\,\kappa}{\log
^{\gamma-1}\left(  K\right)  }+\frac{k_{s}\epsilon_{-}\log\left(  L\right)
}{L-1}+\frac{\epsilon_{-}\kappa}{\epsilon_{+}\log^{\gamma}\left(  L\right)
}\right) \\
&  =\epsilon_{-}^{-1}\log^{-1}\left(  L\right)  \,\left(  1+O\left(  \frac
{1}{\log^{\gamma-1}\left(  L\right)  }\right)  \right)  .
\end{align*}
for $\gamma>1$. Therefore, we have finally%
\begin{align*}
C_{d}\left(  P^{\ast}\right)   &  \geq\frac{1}{K}\log\left(  1+KP^{\ast
}\epsilon_{-}\log\left(  L\right)  \left[  1+O\left(  \frac{1}{\log^{\gamma
-1}\left(  L\right)  }\right)  \right]  ^{-1}\right) \\
&  =\frac{1}{K}\log\left(  1+KP^{\ast}\epsilon_{-}\log\left(  L\right)
\left[  1-O\left(  \frac{1}{\log^{\gamma-1}\left(  L\right)  }\right)
\right]  \right)  .
\end{align*}
A lower bound is obviously
\[
\mathbb{E}_{\mathbf{h}}\left(  h_{\infty}^{-1}\right)  \geq\epsilon_{+}%
^{-1}\log^{-1}\left(  L\right)  \,\left(  1-\frac{\kappa}{\log^{\gamma}\left(
L\right)  }\right)
\]
and we have
\begin{align*}
C_{d}\left(  P^{\ast}\right)   &  \leq\frac{1}{K}\log\left(  1+KP^{\ast
}\epsilon_{+}\log\left(  L\right)  \left(  1-\frac{\kappa}{\log^{\gamma
}\left(  L\right)  }\right)  ^{-1}\right) \\
&  =\frac{1}{K}\log\left(  1+KP^{\ast}\epsilon_{+}\log\left(  L\right)
\left[  1+O\left(  \frac{1}{\log^{\gamma-1}\left(  L\right)  }\right)
\right]  \right)
\end{align*}
and the result follows.
\end{proof}

We can conclude from the proof as follows:

\begin{corollary}
The DLC is finite if there are either

\begin{itemize}
\item  at least two independent channel gains in the frequency domain or

\item  at least two independent path gains in the time domain
\end{itemize}

with Lipschitz continuous marginal distribution function in the neighborhood
of the zero.
\end{corollary}

Interestingly, the DLC compares favorably by the factor $\log\left(  L\right)
$ with the capacity of AWGN in the low SNR regime. Hence, the delay spread
governs the DLC in this region.

Let us make the bound explicit for the Rayleigh fading case. Note that due to
the sum of independent complex path gains the samples of the frequency
response are approximately complex Gaussian distributed anyway (however, the
exact distribution would be very difficult to evaluate) \cite{wunder_03_inf}.

\begin{lemma}
\label{lemma:gauss}Suppose $L$ divides $K$ and under the assumption of complex
Gaussian distributed path gains with uniform PDP the maximum channel gain is
enclosed by the inequalities
\begin{align*}
&  \Pr\left(  \log\left(  L\right)  -\gamma\log\left[  \log\left(  L\right)
\right]  \leq h_{\infty}\leq\log\left(  L\right)  +\gamma\log\left[
\log\left(  L\right)  \right]  \right) \\
&  \geq1-\frac{\kappa}{\log^{\gamma}\left(  L\right)  }%
\end{align*}
where $\gamma>0$ and $\kappa\geq\frac{K}{L-\log^{-\gamma}\left(  L\right)  }$
and $\gamma,L$ sufficiently large such that the bound makes sense.
\end{lemma}

\begin{proof}
see Appendix \ref{proof_of_lemma:gauss}.
\end{proof}

We can apply the result again to the DLC.

\begin{corollary}
\label{corollary:low1}Under the assumption of complex Gaussian distributed
path gains with uniform PDP the low SNR DLC is enclosed by%
\[
\frac{1}{K}\log\left(  1+\kappa_{1}KP^{\ast}\log\left(  L\right)  \right)
\leq C_{d}\left(  P^{\ast}\right)  \leq\frac{1}{K}\log\left(  1+\kappa
_{2}KP^{\ast}\log\left(  L\right)  \right)
\]
where
\[
\kappa_{1}:=\left(  1-\frac{\gamma\log\left[  \log\left(  L\right)  \right]
}{\log\left(  L\right)  }\right)  \left(  1-\frac{1.1\kappa}{\log^{\gamma
-1}\left(  L\right)  }\right)
\]
and%
\[
\kappa_{2}:=\left(  1+\frac{\gamma\log\left[  \log\left(  L\right)  \right]
}{\log\left(  L\right)  }\right)  \left(  1+\frac{1.1\kappa}{\log^{\gamma
}\left(  L\right)  }\right)
\]
for any $\gamma>1,\kappa>0,$ and $L>1$ from Lemma \ref{lemma:gauss}.
\end{corollary}

\begin{proof}
The result follows if the constants in Lemma \ref{lemma:gauss} are used in the
proof of Lemma \ref{lemma:low_snr} and by simple numerical check.
\end{proof}

\begin{remark}
If $L$ does not divide $K$ it seems very difficult to get results that are
asymptotically tight. However, good bounds are easily obtained by observing that
the frequency response cannot arbitrarily overshoot between the samples. In
fact, for the upper bound we can use a grid $K^{\prime}=aL$, say by a factor
$a>1$, and by collecting these samples in $\mathbf{h}^{\left(  a\right)  }$ we
have \cite{wunder_03_sig}%
\[
\left\|  \mathbf{h}\right\|  _{\infty}\leq\left\|  \mathbf{h}^{\left(
a\right)  }\right\|  _{\infty}\cos^{-1}\left(  \frac{\pi}{2a}\right)  .
\]
For the lower bound we set $a=1$ and have%
\[
\left\|  \mathbf{h}\right\|  _{\infty}\geq\left\|  \mathbf{h}^{\left(
1\right)  }\right\|  _{\infty}\cos\left(  \frac{\pi L}{2K}\right)  .
\]
Since we have just derived bounds for $\left\|  \mathbf{h}^{\left(  a\right)
}\right\|  _{\infty},a\geq1$, we can tackle also the general case.
\end{remark}

\begin{figure}[ptbptb]
\begin{center}
\includegraphics[width=10cm]{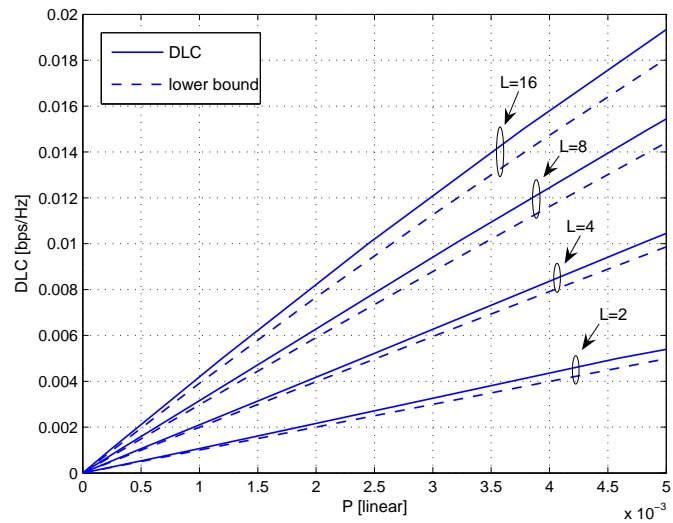}
\end{center}
\caption{Scaling in the low SNR region: The dashed lines indicate the scaling
at low SNR given by $S_{0}=1/K\log_{2}(1+KP^{\ast}\log(K))$}%
\label{fig:dlc1_low}%
\end{figure}

\begin{figure}[ptbptbptbptb]
\begin{center}
\includegraphics[width=10cm]{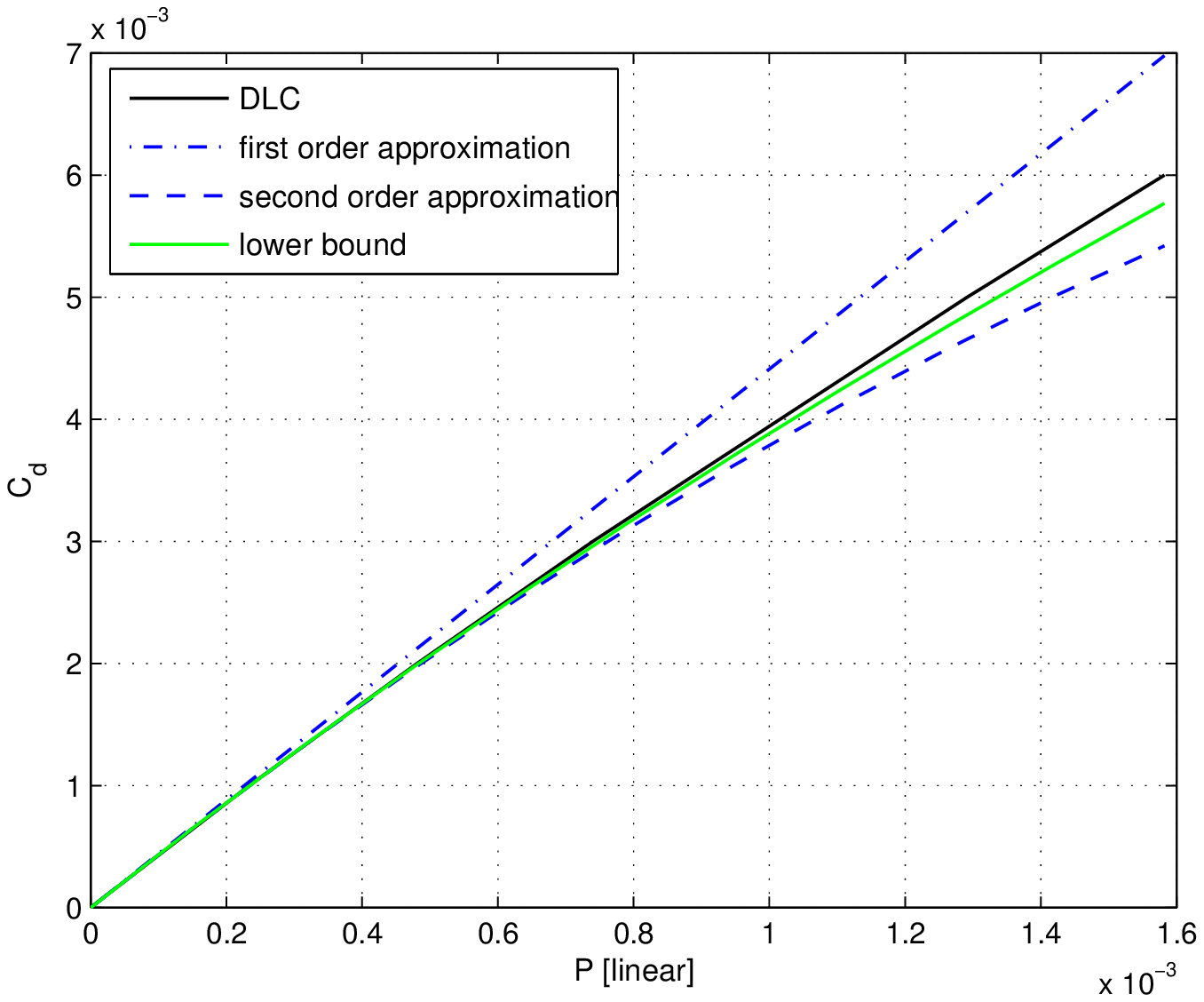}
\end{center}
\caption{Comparison of the DLC with the first and second order approximations
and the lower bound from Fig. \ref{fig:dlc1_low}. Channel with 64 taps delay
spread and system with 64 subcarriers}%
\label{fig:dlc_low_approx_64}%
\end{figure}

\begin{figure}[ptbptbptbptbptb]
\begin{center}
\includegraphics[width=10cm]{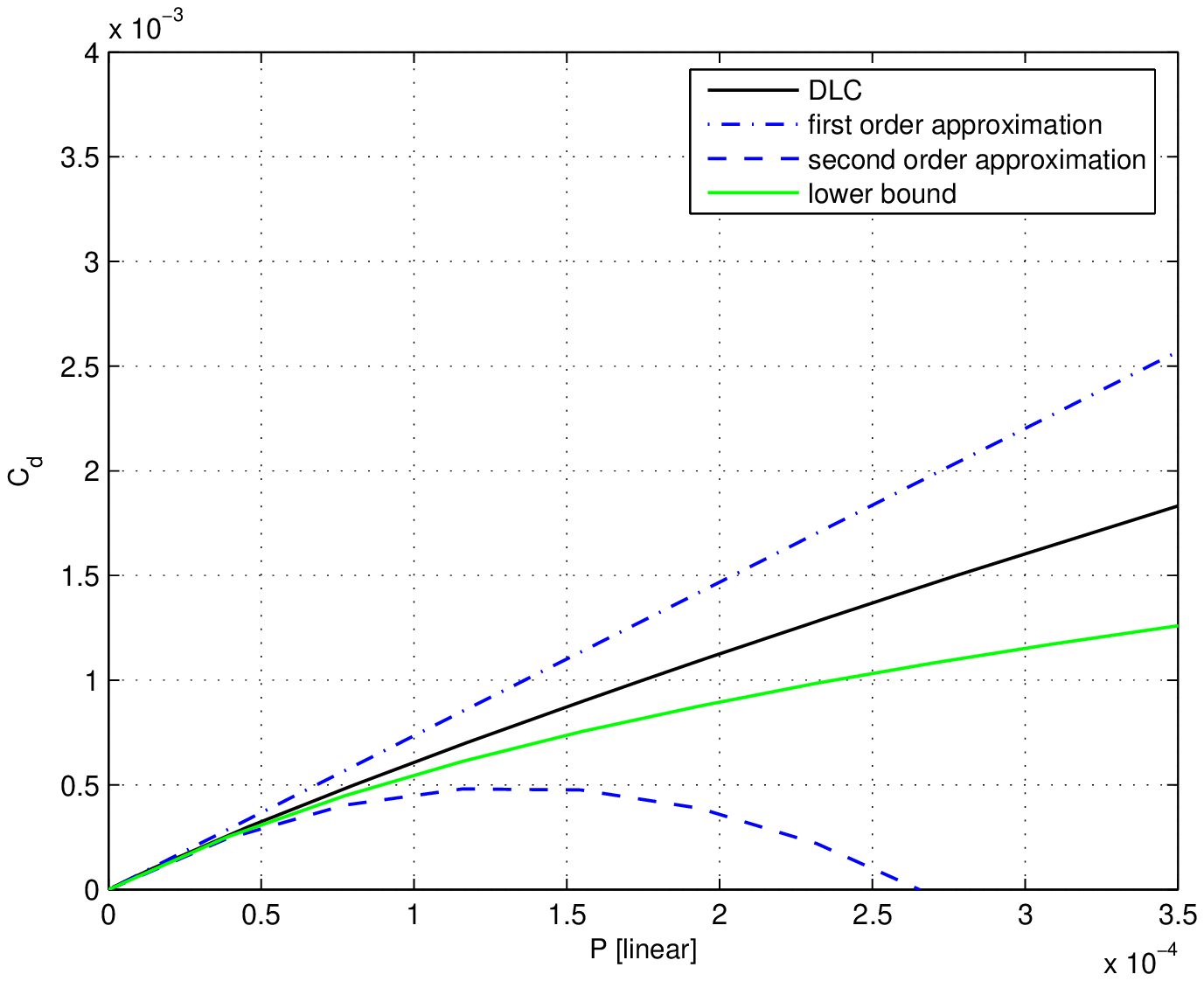}
\end{center}
\caption{Comparison of the DLC with the first and second order approximations
and the lower bound from Fig. \ref{fig:dlc1_low}. Channel with 1024 taps delay
spread and system with 1024 subcarriers}%
\label{fig:dlc_low_approx_1024}%
\end{figure}

An illustration is shown in Fig. \ref{fig:dlc1_low} where we calculate
(\ref{eqn:logf}) for different but low SNR. It is observed that the
approximations are quite accurate for small $L$. The behavior of the OFDM DLC
at low SNR and the corresponding first and second order approximations are
depicted in Figures \ref{fig:dlc_low_approx_64} and
\ref{fig:dlc_low_approx_1024}. It can be seen that the region where the
approximations hold diminishes as the number of degrees of freedom increases.
The bounds can be used to roughly estimate the performance of e.g. a cellular
system at the cell border. Even though we have made not effort to optimize the
bound we found by simulations that for Rayleigh fading with small delay spread
the error is within reasonable span of the optimal curves. However, it is
worth noting that for large ratios of $L$ and $K$ the range where the bound
makes sense becomes increasingly small.

\subsubsection{Impact of power delay profile}

The impact of the PDP has been touched already in Lemma \ref{lemma:low_snr}
proving the ''order-optimality'' of uniform PDP. Let us now investigate the
general case. The following expression can be used to get a bound for
arbitrary PDP.

\begin{proposition}
\label{prop:low1}For sufficiently low SNR an upper bound on the DLC is given
by:
\begin{equation}
C_{d}\left(  P^{\ast}\right)  \leq\frac{1}{K}\log\left(  1+\frac{K^{\left(
1+\alpha\right)  }P^{\ast}}{\mathbb{E}\left(  \left\|  \mathbf{c}\right\|
_{1}^{-1}\right)  }\right)
\end{equation}
Here, $\alpha\in\left(  0,1\right)  $ is a global parameter that can be
numerically optimized. The lower bound is independent of the order of the
elements of the PDP and concave, thus \emph{Schur-concave}.
\end{proposition}

\begin{proof}
The proof follows from Prop. (\ref{prop:low1}) and the inequality chain
$\left\|  \mathbf{h}\right\|  _{1}\geq\left\|  \mathbf{h}\right\|  _{\infty
}\geq K^{-1}\left\|  \mathbf{h}\right\|  _{1}=\left\|  \mathbf{c}\right\|
_{1}$. Since both lower and upper bound are tight only in identifiable special
cases there is some $\alpha\in\left(  0,1\right)  $ that can be numerically
found. The Schur-concavity is obtained from Prop. B.2. in \cite[pp.
287]{marshall_79} since $\left\|  \mathbf{c}\right\|  _{1}^{-1}$ is symmetric
and convex and the expectation $\mathbb{E}\left(  \left\|  \mathbf{c}\right\|
_{1}^{-1}\right)  $ is independent of the ordering of the PDP.
\end{proof}

The Schur-concavity implies that for fixed $L$ and normalized PDP the bound is
larger if the elements of the PDP vector are more ''spread out''. If $L$ is
large the bound approaches the low SNR AWGN capacity time a factor $K^{\alpha
}$ for uniform PDP due to the strong law of large number, i.e. $\left\|
\mathbf{c}\right\|  _{1}\rightarrow1$ almost surely, and hence $\alpha$ is of
order $\log\left[  \log\left(  L\right)  \right]  /\log\left(  L\right)  $.

\subsection{Scaling in high SNR}

Defining the quantity $\overline{h}:=\prod_{k=1}^{K}h_{k}^{-1/K}$\ it was
proved in \cite{biglieri_01_inf} that by using the suboptimal power control
law (\ref{eqn:pclaw}) $C_{d}\left(  P^{\ast}\right)  \geq\log\left(  P^{\ast
}/\mathbb{E}_{\mathbf{h}}\left(  \overline{h}\right)  \right)  $ provided that
$\mathbb{E}_{\mathbf{h}}\left(  \overline{h}\right)  <\infty$, i.e. for
regular fading distributions. We can extend this result to an upper bound
without using any simplifying assumptions on the fading distribution.

\begin{proposition}
\label{prop:high1}Suppose that $\mathbb{E}_{\mathbf{h}}\left(  \overline
{h}\right)  <\infty$. For sufficiently large $P^{\ast}$ the DLC is upper
bounded by
\[
C_{d}\left(  P^{\ast}\right)  \leq\log\left(  \frac{P^{\ast}\left(  1+\frac
{1}{K}\right)  }{\mathbb{E}_{\mathbf{h}}\left(  \overline{h}\right)  }\right)
.
\]
\end{proposition}

\begin{proof}
We can use the following strategy for an upper bound $C_{d}$: fix $\epsilon>0$
and suppose that for any fading state $\mathbf{h}$ we set the values that are
below $\epsilon$ to $\epsilon$. In other words we do not allow ''virtual''
channel gains below $\epsilon$.

Define $h_{k}^{\epsilon}:=\max\left\{  h_{k},\epsilon\right\}  $. Then, using
(\ref{eqn:dlc1}) we have for sufficiently large $P^{\ast}$
\begin{align}
P^{\ast}\geq &  \mathbb{E}_{\mathbf{h}}\left(  e^{C_{d}}\prod_{k=1}^{K}\left(
h_{k}^{\epsilon}\right)  ^{-\frac{1}{K}}\right)  -\frac{1}{K}\sum_{k=1}%
^{K}\mathbb{E}_{\mathbf{h}}\left(  \frac{1}{h_{k}^{\epsilon}}\right)
\label{eqn:all}\\
&  =\mathbb{E}_{\mathbf{h}}\left(  e^{C_{d}}\prod_{k=1}^{K}\left(
h_{k}^{\epsilon}\right)  ^{-\frac{1}{K}}\right)  -\mathbb{E}_{\mathbf{h}%
}\left(  \frac{1}{h_{1}^{\epsilon}}\right)  .\nonumber
\end{align}
Obviously, the second term grows without bound as $\epsilon\rightarrow0$ for
many fading distributions such as Rayleigh fading. Furthermore the growth
depends on $P^{\ast}$. Let us bound this term as follows: we have
\begin{align*}
\mathbb{E}_{\mathbf{h}}\left(  \frac{1}{h_{1}^{\epsilon}}\right)   &
=\int_{0}^{\infty}\frac{1}{h_{1}^{\epsilon}}dF_{\mathbf{h}}\left(
h_{1}\right) \\
&  \leq\epsilon^{-1}.
\end{align*}
Clearly, the term $\epsilon$ is related to $P^{\ast}$. Since the maximum
channel gain is at least $\epsilon$ and the underlying optimal power control
law is water-filling the above equation (\ref{eqn:all}) is certainly true if
\[
P^{\ast}\geq\frac{K}{\epsilon}%
\]
which is a rough estimation. Hence, we obtain
\[
\mathbb{E}_{\mathbf{h}}\left(  \frac{1}{h_{1}^{\epsilon}}\right)  \leq
\frac{P^{\ast}}{K}%
\]
and finally for any $\epsilon>0$
\[
C_{d}\leq\log\left(  \frac{P^{\ast}\left(  1+\frac{1}{K}\right)  }%
{\mathbb{E}_{\mathbf{h}}\left(  \prod_{k=1}^{K}\left(  h_{k}^{\epsilon
}\right)  ^{-\frac{1}{K}}\right)  }\right)  .
\]
Now observe that $\prod_{k=1}^{K}\left(  h_{k}^{\epsilon}\right)  ^{-\frac
{1}{K}}\leq\overline{h}$ and
\[
\lim_{\epsilon\downarrow0}\prod_{k=1}^{K}\left(  h_{k}^{\epsilon}\right)
^{-\frac{1}{K}}=\overline{h}.
\]
Hence, by dominated convergence
\[
\mathbb{E}_{\mathbf{h}}\left(  \prod_{k=1}^{K}\left(  h_{k}^{\epsilon}\right)
^{-\frac{1}{K}}\right)  \rightarrow\mathbb{E}_{\mathbf{h}}\left(  \overline
{h}\right)
\]
provided that
\[
\mathbb{E}_{\mathbf{h}}\left(  \overline{h}\right)  <\infty.
\]
\end{proof}

The proposition states that as long as $\mathbb{E}_{\mathbf{h}}\left(
\overline{h}\right)  <\infty$ the DLC lies in some target corridor determined
by $\mathbb{E}_{\mathbf{h}}\left(  \overline{h}\right)  $. The following
proposition was proved in \cite{biglieri_01_inf} where it is shown that, under
appropriate circumstances, serving all subcarriers equally is sufficient to
achieve the limiting performance.

\begin{proposition}
\label{prop:high2}Suppose that $\mathbb{E}_{\mathbf{h}}\left(  \overline
{h}\right)  <\infty$. If the joint distribution is continuous, then
\[
C_{d}\left(  P^{\ast}\right)  \sim\log\left(  \frac{P^{\ast}}{\mathbb{E}%
_{\mathbf{h}}\left(  \overline{h}\right)  }\right)  .
\]
\end{proposition}

Hence, appealing to Prop. \ref{prop:high1}, and \ref{prop:high2} it suffices
to evaluate the term $\mathbb{E}_{\mathbf{h}}\left(  \overline{h}\right)  $ in
the high SNR regime. The problem of whether or not the high SNR quantity is
non-zero is not touched upon in \cite{biglieri_01_inf}. Let us therefore
derive general conditions under which this is true. As before we could assume
that the fading distribution is Lipschitz continuous on $\mathbb{R}_{+}^{K}$
which is still too restricting though.

\begin{remark}
Note that one is attempted to derive these conditions from the easier low SNR
quantity since the DLC increases with SNR. However, this is in general
misleading since the joint fading distribution might e.g. contain point masses
on the boundary of the positive orthant rendering the high SNR term infinite
while the low SNR term still provides a proper lower bound.
\end{remark}

The following proposition puts a general lower bound on the DLC and will be
used in Lemma (\ref{lema:high_snr}). In order to not overload the formalism we
assume that the joint distribution on a equidistant subset with some distance
$a\in\mathbb{N}$ of $b\in\mathbb{N}$ subcarriers possesses a density as
follows. Define $\mathbf{h}\left(  a,k_{1}\right)  :=\left[  h_{k_{1}%
+0a},h_{k_{1}+a}\ldots,h_{k_{1}+ab}\right]  ^{T},k_{1}\in\left(
1,\ldots,a\right)  $. We define the following class:%
\begin{align*}
\mathcal{F}_{\mathrm{Hi}}\left(  c_{s}\right)   &  :=\left\{  F_{\mathbf{h}%
\left(  a,k_{1}\right)  }\text{ has density }f_{\mathbf{h}\left(
a,k_{1}\right)  }\text{; }\right. \\
&  f_{\mathbf{h}\left(  a,k_{1}\right)  }\text{ and all marginal densities are
bounded by }c_{s}\\
&  \left.  \text{within }1/b\text{-neighborhood for all }k_{1}\in\left(
1,\ldots,a\right)  \right. \\
&  \left.  \mathbf{h}\in\mathbb{R}_{+}^{K}\text{ is generated from
}\mathbf{\tilde{c}}\in\mathbb{C}_{+}^{L}\text{ by means of (\ref{eqn:channel}%
)}\right\}
\end{align*}
Observe that we have not excluded point masses in this definition.

\begin{proposition}
\label{prop:high3}Suppose that the joint fading distribution on this subset
belongs to $\mathcal{F}_{\mathrm{Hi}}\left(  c_{s}\right)  $. Then, for
sufficiently large $P^{\ast}$ the following lower bound holds:
\begin{equation}
C_{d}\left(  P^{\ast}\right)  \geq\log\left(  P^{\ast}\right)  -\log\left(
b\,c_{s}\left(  \frac{b}{b-1}\right)  ^{b}+\left(  1-c_{s}\right)  b\right)
\label{eqn:a_1}%
\end{equation}
If $a$ can be chosen to be $K/2$, i.e. $b=2$ for even $K$ we have:
\begin{equation}
C_{d}\left(  P^{\ast}\right)  \geq\log\left(  P^{\ast}\right)  -\log\left(
6c_{s}+2\right)  \label{eqn:a_2}%
\end{equation}
\end{proposition}

\begin{proof}
see Appendix \ref{proof_of_prop:high3}.
\end{proof}

Clearly, the bound can be improved in case of independent subcarriers which
will be used in Cor. (\ref{corollary:high1}).

\begin{proposition}
\label{prop:high4}Suppose that the joint fading distribution can be written as
$F_{\mathbf{h}\left(  a,k_{1}\right)  }\equiv\prod_{i=1}^{b}f$ for all $k_{1}%
$\footnote{It is straightforward to see (by the structure of the FFT) that if
the fading distribution is generated by a complex path gain distribution of
which the density can be written as $f_{\tilde{c}_{i}}\left(  \tilde{c}%
_{i}\right)  ,\tilde{c}_{i}\in\mathbb{C}$, where $f_{\tilde{c}_{i}}$ is
invariant under complex rotations, then the fading distribution is also
invariant regarding $k_{1}$.}, i.e. subsets of the subcarriers are
independent, and that the marginal channel gain density $f$ is finite
everywhere. Then, for sufficiently large $P^{\ast}$ the lower bounds
(\ref{eqn:a_1}),(\ref{eqn:a_2}) can be improved to give:
\begin{align*}
C_{d}\left(  P^{\ast}\right)   &  \geq\log\left(  P^{\ast}\right)
-b\log\left(  \frac{b}{b-1}\right) \\
&  -\log\left[  \left(  \sum_{i\geq1}f\left(  a_{i}^{-}\right)  -f\left(
b_{i}^{-}\right)  \right)  ^{1/b}-\sum_{i\geq1}a_{i}^{\frac{b-1}{b}}\left(
f\left(  b_{i}^{+}\right)  -f\left(  a_{i}^{+}\right)  \right)  \right]
\end{align*}
Here, $0\leq a_{i}^{-/+}<b_{i}^{-/+}\leq\infty$ are interval boundaries of
$\mathrm{supp}\left(  f\right)  $ such that $f^{\prime}\left(  h\right)
\leq0,h\in\left[  a_{i}^{-},b_{i}^{-}\right]  $ and $f^{\prime}\left(
h\right)  \geq0,h\in\left[  a_{i}^{+},b_{i}^{+}\right]  $. If $b$ is large and
$f^{\prime}\left(  h\right)  \leq0,h\in\mathbb{R}$, with $f\left(  0\right)
=1$ (e.g. Rayleigh fading), then $\left(  b/b-1\right)  ^{b}\rightarrow e^{1}$
and hence $C_{d}\left(  P^{\ast}\right)  \geq\log\left(  P^{\ast}\right)  -1$.
\end{proposition}

\begin{proof}
see Appendix \ref{proof_of_prop:high4}.
\end{proof}

For the next proposition we need explicit the properties of the FFT.

\begin{proposition}
\label{prop:high5}Suppose that $L,K$ are even and that the densities of real
and imaginary parts of the complex path gain distribution fulfill $f_{\Re
e\left(  \tilde{c}_{i}\right)  ,\Im m\left(  \tilde{c}_{i}\right)  }\left(
x\right)  \leq\sqrt{v_{i}}e^{-\alpha\left|  x\right|  ^{2}}$. Then, the
following lower bound holds:%
\[
C_{d}\left(  P^{\ast}\right)  \geq\log\left(  P^{\ast}\right)  -\log\left(
\frac{4\pi^{L}\prod_{k=1}^{L}v_{k}^{1/2}}{L^{L}}\left(  \frac{L}{\alpha
}\right)  ^{L-1}\right)
\]
\end{proposition}

\begin{proof}
see Appendix \ref{proof_of_prop:high5}.
\end{proof}

The latter proposition is universal but tailored to the Rayleigh fading case.
More sophisticated bounding techniques can be obtained from mixing the
techniques of Prop. (\ref{prop:high3}) to Prop. (\ref{prop:high5}) as
discussed in the remark of Appendix \ref{proof_of_prop:high5}. In the
following we assume without loss of generality that the marginal fading
distributions are such that $\mathbb{E}_{\mathbf{h}}\left(  \log\left(
h_{k}\right)  \right)  $ is independent of $k\in\mathcal{K}$ (and finite)
where $K$ is supposed here to be even.

\begin{lemma}
\label{lema:high_snr}Suppose that $\mathbb{E}_{\mathbf{h}}\left(  \overline
{h}\right)  <\infty$. For sufficiently large $P^{\ast}$ the DLC is upper
bounded by
\[
C_{d}\left(  P^{\ast}\right)  \leq\log\left(  P^{\ast}\right)  +H\left(
F_{\mathbf{h}}\right)  +\frac{1}{K}%
\]
where
\[
H\left(  F_{\mathbf{h}}\right)  :=\int_{0}^{\infty}\log\left(  h\right)
\,dF_{h_{1}}\left(  h\right)
\]
and $F_{h_{1}}$ is the marginal fading distribution. Furthermore, suppose that
there is a sequence of fading distributions in $\mathcal{F}_{\mathrm{Hi}%
}\left(  c_{s}\right)  $ such that $K^{-1}\sum_{k=1}^{K}\log\left(
h_{k}\right)  \rightarrow H\left(  F_{\mathbf{h}}\right)  $ in probability.
Then, the bound is asymptotically tight, i.e.
\[
C_{d}\left(  P^{\ast}\right)  \rightarrow\log\left(  P^{\ast}\right)
+H\left(  F_{\mathbf{h}}\right)  .
\]
\end{lemma}

\begin{proof}
Setting
\[
\mathbb{E}_{\mathbf{h}}\left(  \overline{h}\right)  =\mathbb{E}_{\mathbf{h}%
}\left(  \exp\left[  \log\left(  \overline{h}\right)  \right]  \right)
\]
we get by Jensen's inequality
\begin{align*}
&  \mathbb{E}_{\mathbf{h}}\left(  \exp\left[  \log\left(  \overline{h}\right)
\right]  \right)  \geq\exp\left(  \mathbb{E}\left[  \log\left(  \overline
{h}\right)  \right]  \right) \\
&  =\exp\left(  \int_{0}^{\infty}\log\left(  h\right)  \,dF_{h_{1}}\left(
h\right)  \right)
\end{align*}
which is already the desired upper bound provided that $H\left(
F_{\mathbf{h}}\right)  <\infty$ or equivalently $\mathbb{E}_{\mathbf{h}%
}\left(  \overline{h}\right)  <\infty$.

In order to show the tightness of the upper bound define the following random
variable (i.e. partial sums):
\begin{equation}
h^{\left(  K\right)  }:=-\frac{1}{K}\sum_{k=1}^{K}\log\left(  h_{k}\right)  .
\label{eqn:h_K}%
\end{equation}
Suppose that $h^{\left(  K\right)  }\rightarrow H\left(  F_{\mathbf{h}%
}\right)  $ in probability. We have to show that
\begin{align*}
&  E_{\mathbf{h}}\left(  \exp\left[  h^{\left(  K\right)  }\right]  \right) \\
&  \rightarrow\exp\left(  -\int_{0}^{\infty}\log\left(  h\right)  \,dF_{h_{1}%
}\left(  h\right)  \right)  ,K\rightarrow\infty,
\end{align*}
which would follow if $h^{\left(  K\right)  }$ is uniformly bounded in $K$ but
is not true for the situation at hand. Using the set function $\mathbb{I}%
\left\{  \cdot\right\}  $ and writing for some $C>0$%
\begin{align*}
&  \mathbb{E}_{\mathbf{h}}\left(  \exp\left(  h^{\left(  K\right)  }\right)
\right)  =\mathbb{E}_{\mathbf{h}}\left(  \exp\left(  h^{\left(  K\right)
}\right)  \mathbb{I}\left\{  h^{\left(  K\right)  }\leq C\right\}  \right) \\
&  +\mathbb{E}_{\mathbf{h}}\left(  \exp\left(  h^{\left(  K\right)  }\right)
\mathbb{I}\left\{  h^{\left(  K\right)  }>C\right\}  \right) \\
&  \leq\mathbb{E}_{\mathbf{h}}\left(  \min\left\{  \exp\left(  h^{\left(
K\right)  }\right)  ,\exp\left(  C\right)  \right\}  \right) \\
&  +\mathbb{E}_{\mathbf{h}}\left(  \exp\left(  h^{\left(  K\right)  }\right)
\mathbb{I}\left\{  h^{\left(  K\right)  }>C\right\}  \right)
\end{align*}
yields for the first term on the RHS
\begin{align*}
&  \mathbb{E}_{\mathbf{h}}\left(  \min\left\{  \exp\left(  h^{\left(
K\right)  }\right)  ,\exp\left(  C\right)  \right\}  \right) \\
&  \rightarrow\exp\left[  \int_{0}^{\infty}\log\left(  h\right)  \,dF_{h_{1}%
}\left(  h\right)  \right]  ,\quad K\rightarrow\infty,
\end{align*}
since $h^{\left(  K\right)  }\rightarrow H\left(  F_{\mathbf{h}}\right)  $ in
probability and $\min\left\{  \exp\left(  h^{\left(  K\right)  }\right)
,C\right\}  $ is uniformly bounded (and $C$ sufficiently large!). Hence, we
have
\begin{align*}
&  \underset{n\rightarrow\infty}{\lim\sup}\,\mathbb{E}_{\mathbf{h}}\left(
\exp\left(  h^{\left(  K\right)  }\right)  \right) \\
&  \leq H\left(  F_{\mathbf{h}}\right)  +\underset{n\rightarrow\infty}%
{\lim\sup}\,\mathbb{E}_{\mathbf{h}}\left(  \exp\left(  h^{\left(  K\right)
}\right)  \mathbb{I}\left\{  h^{\left(  K\right)  }>C\right\}  \right) \\
&  =H\left(  F_{\mathbf{h}}\right)  +\underset{n\rightarrow\infty}{\lim\sup
}\,\int_{\mathbb{R}_{+}^{K}}\overline{h}\,\mathbb{I}\left\{  h^{\left(
K\right)  }>C\right\}  \,dF_{\mathbf{h}}\left(  \mathbf{h}\right)
\end{align*}
Fix $\epsilon>0$ we have by the inequality shown in (\ref{eqn:lieb}, Appendix
\ref{proof_of_prop:high3})%
\[
\int_{\mathbb{R}_{+}^{K}}\overline{h}\,\mathbb{I}\left\{  h^{\left(  K\right)
}>C\right\}  \,dF_{\mathbf{h}}\left(  \mathbf{h}\right)  \leq\prod_{l=1}%
^{K/2}\int_{\mathbb{R}_{+}^{K}\cap\Omega_{C}}\overline{h}_{l}^{\frac{1}{2}%
}\overline{h}_{l+K/2}^{\frac{1}{2}}\,dF_{\mathbf{h}}\left(  \mathbf{h}%
\right)
\]
where $\Omega_{C}:=\left\{  \mathbf{h}\in\mathbb{R}_{+}^{K}:h^{\left(
K\right)  }>C\right\}  $. Then by geometric mean inequality and a ''sandwich''
argument
\begin{align*}
&  \int_{\mathbb{R}_{+}^{K}}\overline{h}\,\mathbb{I}\left\{  h^{\left(
K\right)  }>C\right\}  \,dF_{\mathbf{h}}\left(  \mathbf{h}\right) \\
&  \leq\frac{2}{K}\sum_{l=1}^{K/2}\left(  \int_{\mathbb{R}_{+}^{K}\cap
\Omega_{C}}\left(  \overline{h}_{l}^{\frac{1}{2}}\overline{h}_{l+K/2}%
^{\frac{1}{2}}\,-\left(  \overline{h}_{l}^{\epsilon}\right)  ^{\frac{1}{2}%
}\left(  \overline{h}_{l+K/2}^{\epsilon}\right)  ^{\frac{1}{2}}\right)
\,dF_{\mathbf{h}}\left(  \mathbf{h}\right)  \right) \\
&  +\frac{2}{K}\sum_{l=1}^{K/2}\left(  \int_{\mathbb{R}_{+}^{K}\cap\Omega_{C}%
}\left(  \overline{h}_{l}^{\epsilon}\right)  ^{\frac{1}{2}}\left(
\overline{h}_{l+K/2}^{\epsilon}\right)  ^{\frac{1}{2}}\,dF_{\mathbf{h}}\left(
\mathbf{h}\right)  \right) \\
&  \leq\epsilon^{-1}\Pr\left(  h^{\left(  K\right)  }>C\right)  +C_{\epsilon}%
\end{align*}
where $\Pr\left(  h^{\left(  K\right)  }>C\right)  \rightarrow0$ for
$K\rightarrow\infty$ ($C$ again sufficiently large). The remaining constant
$C_{\epsilon}>0$ can be made arbitrarily small independent of $K$ since
$F_{\mathbf{h}}\in\mathcal{F}_{\mathrm{Hi}}\left(  c_{s}\right)  $ uniformly.
On the other hand, since
\[
\underset{n\rightarrow\infty}{\lim\inf}\,\mathbb{E}_{\mathbf{h}}\left(
\min\left\{  \exp\left(  h^{\left(  K\right)  }\right)  ,\exp\left(  C\right)
\right\}  \right)  \geq H\left(  F_{\mathbf{h}}\right)
\]
we have the desired result.
\end{proof}

The required convergence in probability follows if:

\begin{itemize}
\item  either, the subcarriers are independent (or a subset),

\item  or the logarithmic channel gains are uncorrelated (or a subset) with
\begin{equation}
\frac{1}{K^{2}}\sum_{k=1}^{K}\mathbb{E}_{\mathbf{h}}\left(  \log^{2}\left(
h_{k}\right)  \right)  \rightarrow0,\quad K\rightarrow\infty. \label{eqn:logh}%
\end{equation}
\end{itemize}

Interestingly, since $H\left(  F_{\mathbf{h}}\right)  <0$ there is always a
loss in capacity compared to AWGN. Furthermore, observe that the second
statement (\ref{eqn:logh}) is substantially weaker than independence. It
suggests that ergodic capacity can be achieved even if the subcarriers are not
independent which is discussed in the next subsection. We can apply this
result to the Rayleigh fading case.

\begin{theorem}
   \label{corollary:high1}
   Under the assumption of complex Gaussian distributed
   path gains, the upper bound
   \[
   C_{d}\left(  P^{\ast}\right)  \leq\log\left(  P^{\ast}\right)  +H\left(
     F_{\mathbf{h}}\right)  +\frac{1}{K}%
   \]
   holds where
   \[
   H\left(  F_{\mathbf{h}}\right)  :=\int_{0}^{\infty}\log\left(  h\right)
   \exp\left(  h\right)  dh\approx-0.58.
   \]
   The bound is asymptotically tight for uniform PDP and sequences $\left(
     L_{n},K_{n}\right)  $ where $L_{n}$ divides $K_{n}$, i.e.%
   \[
   C_{d}\left(  P^{\ast}\right)  \rightarrow\log\left(  P^{\ast}\right)
   +H\left(  F_{\mathbf{h}}\right)  ,\quad n\rightarrow\infty.
   \]
   The convergence speed is given by:
   \[
   C_{d}\left(  P^{\ast}\right)  \geq\log\left(  P^{\ast}\right)  +H\left(
     F_{\mathbf{h}}\right)  +O\left(  \log^{-1}\left(  K\right)  \right)
   \]
   (see eqn. (\ref{eqn:speed}) for constants)
\end{theorem}

\begin{proof}
   The first part of the theorem follows immediately from Lemma
   \ref{lema:high_snr} and the fact that subsets of $L$ subcarriers are
   independent. It remains to provide an explicit expression for the convergence
   speed. Let $h^{\left(  K\right)  }$ be defined as as in (\ref{eqn:h_K}). Then
   by Lemma \ref{lema:high_snr} we have to investigate the following terms
   \begin{align*}
      &  \mathbb{E}_{\mathbf{h}}\left(  \exp\left(  h^{\left(  K\right)  }\right)
      \right) \\
      &  \leq\mathbb{E}_{\mathbf{h}}\left(  \min\left\{  \exp\left(  h^{\left(
                K\right)  }\right)  ,C\right\}  \right)  +\mathbb{E}_{\mathbf{h}}\left(
        \exp\left(  h^{\left(  K\right)  }\right)  \mathbb{I}\left\{  h^{\left(
              K\right)  }>C\right\}  \right)
   \end{align*}
   where $C>H\left(  F_{\mathbf{h}}\right)  $. Defining now the event
   $\mathcal{A}=\left\{  \left|  h^{\left(  K\right)  }-\exp\left(  H\left(
           F_{\mathbf{h}}\right)  \right)  \right|  \leq\epsilon\right\}  $ and its
   complement $\mathcal{A}^{C}$ we have
   \begin{align*}
      \mathbb{E}_{\mathbf{h}}\left(  \exp\left(  h^{\left(  K\right)  }\right)
      \right)   &  \leq\mathbb{E}_{\mathbf{h}}\left(  \min\left\{  \exp\left(
            h^{\left(  K\right)  }\right)  ,C\right\}  \mathbb{I}\left\{  \mathcal{A}%
        \right\}  \right) \\
&  +\mathbb{E}_{\mathbf{h}}\left(  \min\left\{  \exp\left(  h^{\left(
K\right)  }\right)  ,C\right\}  \mathbb{I}\left\{  \mathcal{A}^{C}\right\}
\right) \\
&  +\mathbb{E}_{\mathbf{h}}\left(  \exp\left(  h^{\left(  K\right)  }\right)
\mathbb{I}\left\{  h^{\left(  K\right)  }>C\right\}  \right)
\end{align*}
The first two terms can be bounded as follows: since
\[
\mathbb{E}_{\mathbf{h}}\left(  \min\left\{  \exp\left(  h^{\left(  K\right)
}\right)  ,C\right\}  \mathbb{I}\left\{  \mathcal{A}\right\}  \right)
\leq\exp\left(  H\left(  F_{\mathbf{h}}\right)  \right)  +\epsilon
\]
and
\[
\mathbb{E}_{\mathbf{h}}\left(  \min\left\{  \exp\left(  h^{\left(  K\right)
}\right)  ,C\right\}  \mathbb{I}\left\{  \mathcal{A}^{C}\right\}  \right)
\leq C\Pr\left(  \mathcal{A}^{C}\right)
\]
we need a bound on the probability $\Pr\left(  \mathcal{A}^{C}\right)  $
dependent on $\epsilon$. The probability can be easily upper bounded by
Tschebyscheff's inequality, i.e.
\[
\Pr\left(  \mathcal{A}^{C}\right)  \leq\frac{\sigma^{2}}{K\epsilon^{2}}%
\]
where
\[
\sigma^{2}=\mathbb{E}_{\mathbf{h}}\left(  \left[  \log\left(  h_{1}\right)
-\mathbb{E}_{\mathbf{h}}\left(  \log\left(  h_{1}\right)  \right)  \right]
^{2}\right)
\]
and by choosing some sufficiently slowly converging zero sequence, e.g.
$\epsilon_{K}=O\left(  1/\log\left(  K\right)  \right)  $.

The third term can be upper bounded by observing that:%
\begin{align*}
&  \mathbb{E}_{\mathbf{h}}\left(  \exp\left(  h^{\left(  K\right)  }\right)
\mathbb{I}\left\{  h^{\left(  K\right)  }>C\right\}  \right) \\
&  =\int_{\mathbb{R}_{+}^{K}}\overline{h}\,\mathbb{I}\left\{  h^{\left(
K\right)  }>C\right\}  \,dF_{\mathbf{h}}\left(  \mathbf{h}\right) \\
&  \leq\left(  \int_{\mathbb{R}_{+}^{K}}\left(  \overline{h}\right)
^{2}\,dF_{\mathbf{h}}\left(  \mathbf{h}\right)  \right)  ^{\frac{1}{2}}\left(
\int_{\mathbb{R}_{+}^{K}}\left(  \mathbb{I}\left\{  h^{\left(  K\right)
}>C\right\}  \right)  ^{2}\,dF_{\mathbf{h}}\left(  \mathbf{h}\right)  \right)
^{\frac{1}{2}}\\
&  \leq\sqrt{\exp\left(  1\right)  }\left(  \Pr\left(  \left\{  h^{\left(
K\right)  }>C\right\}  \right)  \right)  ^{\frac{1}{2}}%
\end{align*}
In the last inequality we employed Prop. \ref{prop:high4} for the first
integral. The probability can again be tackled with Tschebyscheff's inequality.
It follows
\[
\Pr\left(  \left\{  h^{\left(  K\right)  }>C\right\}  \right)  \leq
\frac{\sigma^{2}}{K\left(  C-H\left(  F_{\mathbf{h}}\right)  \right)  ^{2}}%
\]
and hence:
\begin{align}
&  C_{d}\left(  P^{\ast}\right) \nonumber\\
&  \geq\log\left(  P^{\ast}\right)  -\left(  \exp\left(  H\left(
F_{\mathbf{h}}\right)  \right)  +\epsilon_{K}+\frac{C\sigma^{2}}{K\epsilon
_{K}^{2}}+\frac{\exp^{1/2}\left(  1\right)  \,\sigma}{K^{1/2}\left(
C-H\left(  F_{\mathbf{h}}\right)  \right)  }\right) \nonumber\\
&  \geq\log\left(  P^{\ast}\right)  -\left(  \exp\left(  H\left(
F_{\mathbf{h}}\right)  \right)  +\frac{1}{\log\left(  K\right)  }%
+\frac{C\sigma^{2}\log^{2}\left(  K\right)  }{K}+\frac{\exp^{1/2}\left(
1\right)  \,\sigma}{K^{1/2}\left(  C-H\left(  F_{\mathbf{h}}\right)  \right)
}\right)  \label{eqn:speed}%
\end{align}
\end{proof}

\begin{figure}[ptbptbptb]
\begin{center}
\includegraphics
[width=10cm]{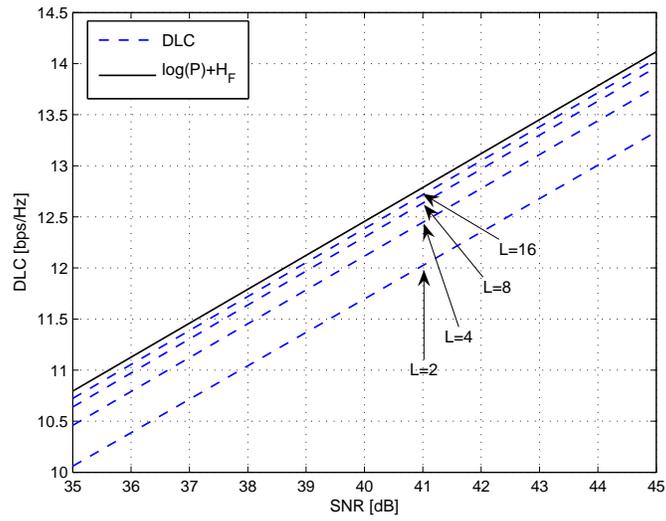}
\end{center}
\caption{Scaling in the high SNR region: The black line indicates the scaling
at high SNR given by $S=\log\left(  P^{\ast}\right)  +H_{F}$ (for Rayleigh
fading). The dashed lines give the OFDM DLC for $L=K=2,4,8,16$.}%
\label{fig:dlc1_high}%
\end{figure}

An illustration is shown in Figure \ref{fig:dlc1_high}.

\subsubsection{Impact of power delay profile}

The impact of the PDP has been touched already in Cor. \ref{corollary:high1}
showing its asymptotic optimality. Similar to the low SNR regime we have an
upper bound describing generally the impact of PDP.

\begin{proposition}
For sufficiently large $P^{\ast}$ (and $K$) the following upper bound holds:%
\[
C_{d}\left(  P^{\ast}\right)  \leq\log\left(  \frac{P^{\ast}}{\mathbb{E}%
_{\mathbf{h}}\left(  \left\|  \mathbf{c}\right\|  _{1}^{-1}\right)  }\right)
\]
The bound is independent of the order of the elements of the PDP and concave
(thus Schur-concave).
\end{proposition}

\begin{proof}
The proof follows immediately from the geometric mean inequality, i.e. for any
$\mathbf{h}$ we have
\[
\overline{h}\geq\frac{K}{\left\|  \mathbf{h}\right\|  _{1}}=\frac{1}{\left\|
\mathbf{c}\right\|  _{1}}.
\]
Taking expectations on both sides in combination with Prop. B.2. in
\cite{marshall_79} yields the desired result.
\end{proof}

If $L$ is large with uniform PDP the bound approaches the high SNR AWGN
capacity and is thus too optimistic in general.

\subsection{Convergence to the ergodic capacity}

Let us now treat the ergodic case. The ergodic capacity is given by
\cite{biglieri_01_inf}:%
\begin{align*}
C_{e}  &  =\mathbb{E}_{\mathbf{h}}\left(  \max\left\{  \log\left(  \xi
h_{1}\right)  ,0\right\}  \right)  \quad\text{where }\xi\text{ such that}\\
P^{\ast}  &  =\mathbb{E}_{\mathbf{h}}\left(  \max\left\{  \xi-h_{1}%
^{-1},0\right\}  \right)
\end{align*}

Note that for low SNR the first order term of the DLC is not bounded with
respect to $L$. Hence, the ergodic capacity has the same property which is in
accordance with results in \cite{Ver:2002}. A similar convergence can be shown
for the high SNR case.

\begin{corollary}
Under the assumptions of Lemma \ref{lema:high_snr} the DLC converges to the
ergodic capacity as $K\rightarrow\infty$.
\end{corollary}

\begin{proof}
We only have to show that $C_{e}$ scales as $\log\left(  P^{\ast}\right)
+H_{F}$ as $P^{\ast}\rightarrow\infty$. We can again use a truncation
argument. Let $h_{1}^{\epsilon}:=\max\left\{  h_{1},\epsilon\right\}  $. Then,
we have for sufficiently large $P^{\ast}$%
\[
\xi-\mathbb{E}_{\mathbf{h}}\left(  \left(  h_{1}^{\epsilon}\right)
^{-1}\right)  =P^{\ast}%
\]
and
\begin{align*}
C_{e}  &  =\mathbb{E}_{\mathbf{h}}\left(  \log\left(  \xi h_{1}^{\epsilon
}\right)  \right) \\
&  =\mathbb{E}_{\mathbf{h}}\left(  \log\left(  \left(  \mathbb{E}_{\mathbf{h}%
}\left(  \left(  h_{1}^{\epsilon}\right)  ^{-1}\right)  +P^{\ast}\right)
h_{1}^{\epsilon}\right)  \right) \\
&  =\log\left(  P^{\ast}\right)  +\mathbb{E}_{\mathbf{h}}\left(  \log\left(
h_{1}^{\epsilon}\right)  \right)  +\log\left(  \mathbb{E}_{\mathbf{h}}\left(
\left(  h_{1}^{\epsilon}\right)  ^{-1}\right)  \left(  P^{\ast}\right)
^{-1}+1\right)
\end{align*}
Now, again by a bounded convergence argument%
\[
\lim_{\epsilon\downarrow0}\mathbb{E}_{\mathbf{h}}\left(  \log\left(
h_{1}^{\epsilon}\right)  \right)  =\mathbb{E}_{\mathbf{h}}\left(  \log\left(
h_{1}\right)  \right)
\]
and
\[
\frac{C_{e}}{\log\left(  P^{\ast}\right)  +H_{F}}\rightarrow1,\quad P^{\ast
}\rightarrow\infty,
\]
which completes the proof.
\end{proof}

We have the following coding implications:

%
%
%
%
%
%
%
%
%
%
%
%
%
%
%
%
%
%
%
%
%
%
%
%
%
%
%
%
%

\section{OFDM broadcast channel DLC region}

\label{sec:mu}

The delay-limited region is defined as follows:

\begin{definition}
\label{def:dlc}A rate vector $\mathbf{R}^{\ast}$ lies in the DLC region
$\mathcal{C}_{\mathrm{DL}}\left(  P^{\ast}\right)  $ with sum power
constraints $P^{\ast}$ constraints if and only if for any fading state
$\mathbf{h}$ there is $P^{\prime}$ solving
\[
\min P\text{\quad s.t.\quad}R^{\ast}\in\mathcal{C}_{\mathrm{BC}}\left(
\mathbf{h},P\right)
\]
and
\[
\mathbb{E}_{\mathbf{h}}\left(  P^{\prime}\right)  \leq P^{\ast}%
\]
Furthermore, $R^{\ast}$ is on the boundary $\mathcal{B}_{\mathrm{DL}}\left(
P^{\ast}\right)  $ if and only if
\[
\mathbb{E}_{\mathbf{h}}\left(  P^{\prime}\right)  =P^{\ast}.
\]
\end{definition}

To evaluate the OFDM delay-limited region turns out to be very difficult. This
is due to the fact, that we have only an \emph{implicit} characterization for
the DLC. This means, that we can check for every rate vector $\mathbf{R}$,
whether it lies inside the DLC region or not, simply by solving the dual
problem for each $\mathbf{h}$. To evaluate the expectation necessary to check
the average power condition can be done by Monte-Carlo runs. However, it is
very difficult to determine all rate vectors, which can be achieved with a
fixed sum power constraint $P^{\ast}$.

Nevertheless and although computationally demanding, the OFDM-DLC region can
be calculated up to any desired finite accuracy. To this end we restate the
Algorithm \ref{alg:min_sum_power} from \cite{michel_05_allerton} and define
\begin{align}
n_{m,k}:=\log\Bigg\{e^{\sum\limits_{n>m}R_{\pi_{k}(n),k}}\Big[\frac{\sigma
^{2}}{|h_{\pi_{k}(m),k}|^{2}}  &  +\sum_{j=1}^{m-1}\frac{\sigma^{2}}%
{|h_{\pi_{k}(j),k}|^{2}}\left(  e^{R_{\pi_{k}(j),k}}-1\right)  e^{\sum
\limits_{n=j+1}^{m-1}R_{\pi_{k}(n),k}}\Big]\Bigg\}^{-1}\label{noise}\\
R_{\pi_{k}(m),k}  &  =\Big[\log(\mu_{m})+n_{m,k}\Big]^{+}.
\label{water_filling}%
\end{align}
Then Algorithm \ref{alg:min_sum_power} yields the minimum sum power necessary
to support a set of rates $\mathbf{R}$.

\begin{myalgorithm}
\caption{Iterative "Rate Water-Filling"}\label{alg:min_sum_power}

\begin{myalgo}
\label{rate_waterfilling} \STATE Set $R_{m,k}=0\quad\forall m\in\mathcal{M},\;k=1,...,K$

\WHILE{desired accuracy is not reached}\FOR{$m=1$ to $M$}

\STATE{\bf(1)} Compute the coefficients $n_{m,k}$ (\ref{noise}) for user $m$

\STATE{\bf(2)} Do water-filling with respect to the rates $R_{m,k}$ for user
$m$ according to equation (\ref{water_filling})

\ENDFOR
\ENDWHILE
\end{myalgo}
\end{myalgorithm}

To evaluate the region $\mathcal{C}_{\mathrm{DL}}\left(  P^{\ast}\right)  $,
first the single user DLC rates have to be calculated for all $m$. This is
done by the evaluation of (\ref{eqn:dlc1}) for fixed $\mathcal{C}%
_{\mathrm{DL}}\left(  P^{\ast}\right)  $ and bisection, since $\mathcal{C}%
_{\mathrm{DL}}\left(  P^{\ast}\right)  $ is monotone in $P^{\ast}$. Due to the
convexity of $\mathcal{C}_{\mathrm{DL}}\left(  P^{\ast}\right)  $, any convex
combination $\mathbf{R}_{\mathrm{int}}$ must lie inside $\mathcal{C}%
_{\mathrm{DL}}\left(  P^{\ast}\right)  $. On the other hand, the single user
rates $R_{m}$ are a component-wise upper bound for all other rate vectors.
Since the necessary power $P\left(  \alpha\mathbf{R}_{\mathrm{int}}\right)  $
is monotone in $\alpha$, simple bisection can determine the boundary of the
region for each angle. For any new points on the boundary, the refinement
procedure can be repeated until the desired number of points defining the
border is obtained. This procedure is summarized in Algorithm
\ref{alg:dlc_alg}. Note, that if the distribution and number of taps is the
same for all users, the region is symmetric and can be constructed by
mirroring one calculated sector.

\begin{myalgorithm}
\caption{OFDM DLC Region Algorithm}\label{alg:dlc_alg}

\begin{myalgo}
\STATE{\bf(1)}\ Determine single user DLC (axis points) by evaluation of
single user DLC (see below) and bisection

\WHILE{desired accuracy not reached}

\STATE{\bf(2)}\ For any two neighboring vectors $\mathbf{R}_{1}\in
\mathcal{B}_{\mathrm{DL}}\left(  P^{\ast}\right)  $ and $\mathbf{R}_{2}
\in\mathcal{B}_{\mathrm{DL}}\left(  P^{\ast}\right)  $ on the boundary
calculate interpolated vector $\mathbf{R}_{int}=1/2\left(  \mathbf{R}%
_{1}+\mathbf{R}_{2}\right)  $

\STATE{\bf(3)}\ Adjust $\alpha>1$ by bisection using Alg.
\ref{alg:min_sum_power}\ such that $\alpha\mathbf{R}_{\mathrm{int}}%
\in\mathcal{B}_{\mathrm{DL}}\left(  P^{\ast}\right)  $

\ENDWHILE
\end{myalgo}
\end{myalgorithm}

%
%
%
%
%
%
%
%
%
%
%
%
%
%
%
%
%
%
%
%
%
%
%
%
%
%
%
%
%

\section{OFDMA achievable delay limited rate region}

\label{sec:ofdma}

Compared to the single user case the multiuser DLC region is more difficult to
analyze. In order to get some insight in this case we derive simple resource
allocation schemes based on OFDMA and rate water-filling. To this end, we
assume independence of the subcarriers achieved by complex Gaussian
distributed path gains and uniform PDP. It is worth noting that this can be
imagined as that we take only $L$ independent frequency samples and assume
that the other value are approximately equal in the neighborhood of the $L$
subcarriers. Although seeming quite complicated, the following lemma yields a
simple lower bound on the OFDM DLC region, implying an OFDMA strategy.

\begin{lemma}
\label{lemma:dlc}Let $\mathbf{i}=\left(  i_{1}\ldots,,i_{K}\right)  \in\left[
1,M\right]  ^{K}$ be a multi-index and let the set $\mathcal{K}_{m}%
^{\mathbf{i}}$ count the number of user indices in $\mathbf{i}$ equal to $m$.
Let $\mathcal{I}_{s}\subset\left[  1,M\right]  ^{K}$ be the subset that
contains these multi-indices, where all users occur in the multi-index
$\mathbf{i}$ at least $s$ times, i.e.
\[
\mathcal{I}_{s}=\left\{  \mathbf{i}:\mathcal{K}_{m}^{\mathbf{i}}\geq
s,m\in\mathcal{M}\right\}  .
\]
Then the average required power $P^{\ast}$ to support any rate vector
$\mathbf{R}$ in each fading state is upper bounded by
\[
P^{\ast}\leq\sum_{\mathbf{i}\in\mathcal{I}_{s}}\sum_{m=1}^{M}\sum
_{\substack{p=1\\k\left[  p\right]  \in\mathcal{K}_{m}^{\mathbf{i}}}}^{\left|
\mathcal{K}_{m}^{\mathbf{i}}\right|  }\frac{e^{R_{m,k\left[  p\right]
}^{\mathbf{i}}}-1}{M^{K}\,\zeta_{m,p}^{-1}}+\left(  M^{K}-\left|
\mathcal{I}_{s}\right|  \right)  \sum_{m=1}^{M}\sum_{p=1}^{\left\lfloor
K/M\right\rfloor }\frac{\left(  e^{\bar{R}_{m,k[p]}}-1\right)  }{M^{K}%
\,\theta_{m,p}^{-1}}%
\]
where $R_{m,k\left[  p\right]  }^{\mathbf{i}}$ is the solution to
\[
\left[  \frac{e^{R_{m,k\left[  p\right]  }^{\mathbf{i}}}}{\zeta_{m,p}^{-1}%
}-\lambda\right]  ^{-}=0,\quad\sum_{p=1}^{\left|  \mathcal{K}_{m}^{\mathbf{i}%
}\right|  }R_{m,k\left[  p\right]  }^{\mathbf{i}}=KR_{m}%
\]
and
\begin{equation}
\zeta_{m,p}=\int_{0}^{\infty}\frac{1}{x}\,dF_{h_{k\left[  p\right]  }^{\prime
}}\left(  x\right)  \label{eqn:zetas}%
\end{equation}
with $h_{k\left[  p\right]  }^{\prime}$ being the $p$-th order statistic of
$\left|  \mathcal{K}_{m}^{\mathbf{i}}\right|  $ random variables $h^{\prime}$
with cumulative density function
\[
F_{h^{\prime}}\left(  x\right)  =\left(  1-e^{-x}\right)  ^{M}.
\]
$\bar{R}_{m,k\left[  p\right]  }$ is the solution to
\[
\left[  \frac{e^{\bar{R}_{m,k\left[  p\right]  }}}{\theta_{m,p}^{-1}}%
-\bar{\lambda}\right]  ^{-}=0,\quad\sum_{p=1}^{\left\lfloor K/M\right\rfloor
}\bar{R}_{m,k\left[  p\right]  }=KR_{m}%
\]
and
\[
\theta_{m,p}=\int_{0}^{\infty}\frac{1}{x}\,dF_{h_{k\left[  p\right]  }%
^{\prime\prime}}\left(  x\right)
\]
with $h_{k\left[  p\right]  }^{\prime\prime}$ being the $p$-th order statistic
of $\left\lfloor \frac{K}{M}\right\rfloor $ random variables $h^{\prime\prime
}$ with cumulative density function
\[
F_{h^{\prime\prime}}\left(  x\right)  =1-M\int_{x}^{\infty}\left(
1-e^{-x^{\prime}}\right)  ^{M-2}\left(  e^{x}-e^{-x^{\prime}}\right)
e^{-x^{\prime}}\,dx^{\prime}.
\]
\end{lemma}

\begin{proof}
The basic idea is to distinguish between two cases: The case, where each user
has the best channel at least on $s$ subcarriers and the case where at least
one user has on less than $s$ of the $K$ subcarriers the best channel. With
the multi-index $\mathbf{i}=\left(  i_{1}\ldots,,i_{K}\right)  \in\left[
1,M\right]  ^{K}$ define the event
\[
\mathcal{H}_{\mathbf{i}}:=\left\{  \omega:h_{i_{1},1}\left(  \omega\right)
>\left.  h_{l,1}\left(  \omega\right)  \right|  _{l\neq i_{1}},\ldots
,h_{i_{K},1}>\left.  h_{l,K}\left(  \omega\right)  \right|  _{l\neq i_{K}%
}\right\}  .
\]
Note that by the absolute continuous fading distribution, we have
\[
\sum_{\mathbf{i}\in\mathcal{I}}\Pr\left(  \mathcal{H}_{\mathbf{i}}\right)  =1
\]
since the remaining events occur with zero probability. Thus, we can express
the average power $P^{\ast}$ by
\[
P^{\ast}=\mathbb{E}_{\mathbf{h}}\left(  P^{\prime}\right)  =\frac{1}{M^{K}%
}\sum_{\mathbf{i}\in\mathcal{I}_{s}}\mathbb{E}_{\mathbf{h}}\left(  \left.
P^{\prime}\right|  \mathcal{H}_{\mathbf{i}}\right)  +\frac{1}{M^{K}}%
\sum_{\mathbf{i}\notin\mathcal{I}_{s}}\mathbb{E}_{\mathbf{h}}\left(  \left.
P^{\prime}\right|  \mathcal{H}_{\mathbf{i}}\right)
\]
where $\mathcal{I}_{s}\subset\left[  1,M\right]  ^{K}$ is the subset that
contains the elements where all users occur in the multi-index $\mathbf{i}$ at
least $s$ times. Let $\mathcal{K}_{m}^{\mathbf{i}}$ be the set that counts the
number of user indices in $\mathbf{i}$\ equal to $m$. Fixing $\mathbf{i}$ and
$m$ and ordering the values according to%
\[
h_{m,k\left[  \left|  \mathcal{K}_{m}^{\mathbf{i}}\right|  \right]  }%
\geq\ldots\geq h_{m,k\left[  1\right]  },\quad k\left[  p\right]
\in\mathcal{K}_{m}^{\mathbf{i}}%
\]
the first term on the right hand side is bounded by
\begin{equation}
\sum_{\mathbf{i}\in\mathcal{I}_{s}}\mathbb{E}_{\mathbf{h}}\left(  \left.
P^{\prime}\right|  \mathcal{H}_{\mathbf{i}}\right)  \leq\sum_{\mathbf{i}%
\in\mathcal{I}_{s}}\sum_{m=1}^{M}\sum_{\substack{p=1\\k\left[  p\right]
\in\mathcal{K}_{m}^{\mathbf{i}}}}^{\left|  \mathcal{K}_{m}^{\mathbf{i}%
}\right|  }\mathbb{E}_{\mathbf{h}}\left(  \left.  \frac{e^{R_{m,k\left[
p\right]  }^{\mathbf{i}}}-1}{M^{K}\,h_{m,k\left[  p\right]  }}\right|
\mathcal{H}_{\mathbf{i}}\right)  \label{eq:first_term}%
\end{equation}
with $R_{m,k\left[  p\right]  }^{\mathbf{i}}$ such that the required rates are
supported: Since the expectation on the RHS is independent of the actual
referred subindex in the multi-index $\mathbf{i}$ and depends only on the
number of emerging entries of user $m$ in $\mathbf{i}$ counted by the set
$\mathcal{K}_{m}^{\mathbf{i}}$ we can replace the RHS and rewrite the
inequality in (\ref{eq:first_term}) as
\begin{align*}
\sum_{\mathbf{i}\in\mathcal{I}_{s}}\mathbb{E}_{\mathbf{h}}\left(  \left.
P^{\prime}\right|  \mathcal{H}_{\mathbf{i}}\right)   &  \leq\sum
_{\mathbf{i}\in\mathcal{I}_{s}}\sum_{m=1}^{M}\sum_{\substack{p=1\\k\left[
p\right]  \in\mathcal{K}_{m}^{\mathbf{i}}}}^{\left|  \mathcal{K}%
_{m}^{\mathbf{i}}\right|  }\frac{e^{R_{m,k\left[  p\right]  }^{\mathbf{i}}}%
-1}{M^{K}}\mathbb{E}_{\mathbf{h}}\left(  \frac{1}{h_{m,k\left[  p\right]  }%
}\right) \\
&  \leq\sum_{\mathbf{i}\in\mathcal{I}_{s}}\sum_{m=1}^{M}\sum_{p=1}^{\left|
\mathcal{K}_{m}^{\mathbf{i}}\right|  }\frac{e^{R_{m,k\left[  p\right]
}^{\mathbf{i}}}-1}{M^{K}\,\zeta_{m,p}^{-1}}%
\end{align*}
where $\zeta_{m,p}$ is the expectation of the $p$-th inverse channel
coefficient. The distribution of the $p$-th order of the best channels
$h_{k}^{\max}$ for some $k$ is independent of $k$ and given in equation
(\ref{eqn:order_density}). The cdf and pdf $f\left(  x\right)  $ of
$h_{k}^{\max}$ in turn can also be derived by the order statistic from
(\ref{eqn:order_density}) and be expressed as
\begin{align}
F^{\left(  c\right)  }\left(  x\right)   &  =\frac{\Pr\left(  h_{1,1}%
>x,h_{1,1}>h_{l,1,l\neq1}\right)  }{\Pr\left(  h_{1,1}>h_{l,1,l\neq1}\right)
}\nonumber\\
&  =M\int_{x}^{\infty}\left(  1-e^{-x^{\prime}}\right)  ^{M-1}e^{-x^{\prime}%
}\,dx^{\prime}. \label{eq:cdf}%
\end{align}
and
\begin{equation}
f\left(  x\right)  =-\frac{dF^{c}\left(  x\right)  }{dx}=M\left(
1-e^{-x}\right)  ^{M-1}e^{-x} \label{eq:pdf}%
\end{equation}
Substituting (\ref{eq:cdf}) and (\ref{eq:pdf}) into (\ref{eqn:order_density})
yields the desired densities.

For the second term we need to carry out a different strategy. For all cases
represented in the complementary set $\mathcal{\bar{I}}_{s}$ at least one user
occurs in the multi-index $\mathbf{i}$ less than $s$ times and hence has the
best channel on less than $s$ subcarriers. For the case $s=1$, the strategy of
the previous term can not even guarantee his delay limited rate requirement.
Alternatively, we simply divide the set of subcarriers in $M$ sub-bands where
to each user $\left\lfloor K/M\right\rfloor $ subcarriers are allocated and do
rate water-filling as done for the first term and take the best out of this
set. Hence the second term is upper bounded by
\begin{align}
\sum_{\mathbf{i}\in\mathcal{\bar{I}}_{s}}\mathbb{E}_{\mathbf{h}}\left(
\left.  P^{\prime}\right|  \mathcal{H}_{\mathbf{i}}\right)   &  \leq
\sum_{\mathbf{i}\in\mathcal{\bar{I}}_{s}}\mathbb{E}_{\mathbf{h}}\left(
\left.  \sum_{m=1}^{M}\sum_{p=1}^{\left\lfloor K/M\right\rfloor }\frac
{e^{\bar{R}_{m,k\left[  p\right]  }}-1}{h_{m,k\left[  p\right]  }}\right|
\mathcal{H}_{\mathbf{i}}\right) \nonumber\\
&  \leq\sum_{\mathbf{i}\in\mathcal{\bar{I}}_{s}}\sum_{m=1}^{M}\sum
_{p=1}^{\left\lfloor K/M\right\rfloor }\left(  e^{\bar{R}_{m,k\left[
p\right]  }}-1\right)  \mathbb{E}_{\mathbf{h}}\left(  \left.  \frac
{1}{h_{m,k\left[  p\right]  }}\right|  \mathcal{H}_{\mathbf{i}}^{m}\right)
\label{eqn:term2}%
\end{align}
where the set $\mathcal{H}_{\mathbf{i}}^{m}$ is defined as
\begin{equation}
\mathcal{H}_{\mathbf{i}}^{m}:=\left\{  \omega:h_{i_{k_{m}^{l}},k_{m}^{l}%
}\left(  \omega\right)  >\left.  h_{m,k_{m}^{l}}\left(  \omega\right)
\right|  _{m\neq i_{k_{m}^{l}}},\ldots,h_{i_{k_{m}^{u}},k_{m}^{u}}\left(
\omega\right)  >\left.  h_{m,k_{m}^{u}}\left(  \omega\right)  \right|  _{m\neq
i_{k_{m}^{u}}}\right\}  .
\end{equation}
Note, that the second inequality stems from the fact, that the expectation is
conditioned on the set $\mathcal{H}_{\mathbf{i}}^{m}$, assuming that user $m$
has not the best channel on any of his subcarriers.

Since all subcarriers are independent, we define for each subcarrier the
following conditioned probability and get after some manipulations
\begin{align}
F_{h^{\prime\prime}}\left(  x\right)   &  =1-\Pr\left(  \left.  h_{m,k}%
>x\right|  h_{n,k}>h_{m,k},n\neq m\right) \nonumber\\
&  =1-\frac{\Pr\left(  h_{1,1}>x,h_{2,1}>h_{l,1,l\neq2}\right)  }{\Pr\left(
h_{2,1}>h_{l,1,l\neq2}\right)  }\label{eq:cdf_term2}\\
&  =1-M\int_{x}^{\infty}\left(  1-e^{-x^{\prime}}\right)  ^{M-2}\left(
e^{x}-e^{-x^{\prime}}\right)  f\left(  x^{\prime}\right)  \,dx^{\prime
}.\nonumber
\end{align}
Thus, with (\ref{eqn:order_density}) we can express the condition expectation
as
\[
\mathbb{E}_{\mathbf{h}}\left(  \left.  \frac{1}{h_{m,k\left[  p\right]  }%
}\right|  \mathcal{H}_{\mathbf{i}}^{m}\right)  =\int_{0}^{\infty}\frac{1}%
{x}\,dF_{h_{k\left[  p\right]  }^{\prime\prime}}\left(  x\right)
\]
leading to (\ref{eqn:zetas}). Since the addends do not depend on the index
$\mathbf{i}$, the first sum in (\ref{eqn:term2}) can be substituted by the
factor $\left|  \mathcal{\bar{I}}_{s}\right|  =M^{K}-\left|  \mathcal{I}%
_{s}\right|  $ leading to
\[
\sum_{\mathbf{i}\in\mathcal{\bar{I}}_{s}}\mathbb{E}_{\mathbf{h}}\left(
\left.  P^{\prime}\right|  \mathcal{H}_{\mathbf{i}}\right)  \leq\left(
M^{K}-\left|  \mathcal{I}_{s}\right|  \right)  \sum_{m=1}^{M}\sum
_{p=1}^{\left\lfloor K/M\right\rfloor }\frac{\left(  e^{\bar{R}_{m,k\left[
p\right]  }}-1\right)  }{M^{K}\,\theta_{m,p}^{-1}}.
\]
This concludes the proof.
\end{proof}

Note, that for the case $M=2$, the expression for the cdf in
(\ref{eq:cdf_term2}) simplifies to $F_{h^{\prime\prime}}\left(  x\right)
=1-e^{-2x}$. Instead of partitioning the subcarriers equally among the users,
it is possible to share them in any other relation. Then the second sum of the
second term in (\ref{lemma:dlc}) is not has not $\left\lfloor K/M\right\rfloor
$ addends but $K_{m}$ addends for each user with $\sum_{m=1}^{M}K_{m}=K$. So
it is especially reasonable to share the subcarriers proportional to the users
rate requirements such that $K_{m}=R_{m}/\sum_{m=1}^{M}R_{m}$.

\begin{figure}[ptbptbptbptbptbptbptb]
\begin{center}
\includegraphics[width=10cm]{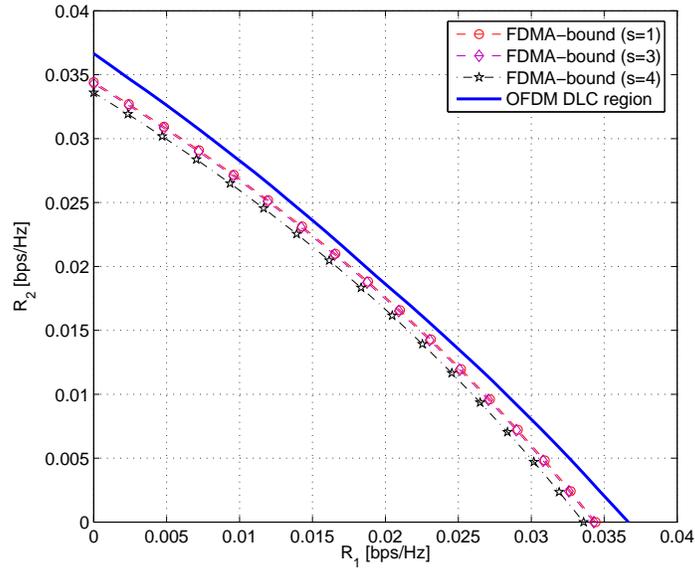}
\end{center}
\caption{OFDM DLC region for 16 subcarriers at -20 dB. The lower bound is
shown for different values of the parameter $s$. The lower bound degrades for
increasing values of $s$.}%
\label{fig:dlc_low}%
\end{figure}

\begin{figure}[ptbptbptbptbptbptbptbptb]
\begin{center}
\includegraphics[width=10cm]{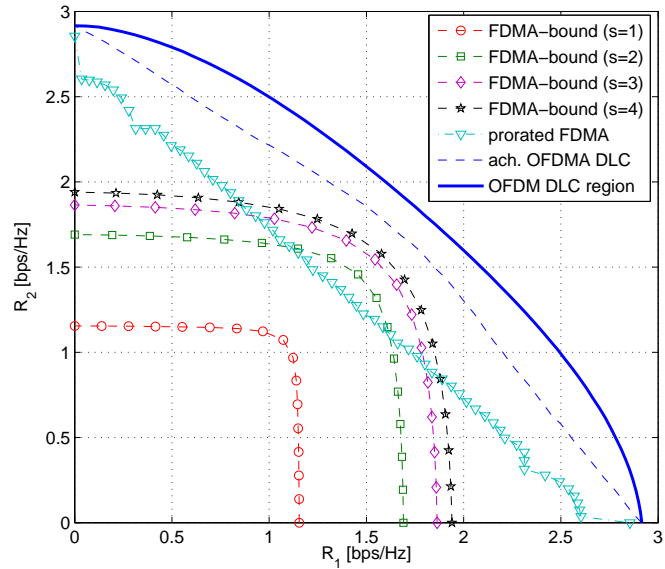}
\end{center}
\caption{OFDM DLC region for 16 subcarriers at 10 dB. The lower bound is shown
for values of the parameter $s=1,...,4$. Further, the prorated FDMA region and
an achievable OFDMA DLC region are depicted.}%
\label{fig:dlc_high}%
\end{figure}

The bounds from the previous section are illustrated in Fig.
\ref{fig:dlc_high} and Fig. \ref{fig:dlc_low}. The bounds are shown for
different values of $s$, changing the relation between term 1 and term 2 in
(\ref{lemma:dlc}). Fig. \ref{fig:dlc_low} depicts the low SNR case. It can be
seen that the lower bound nearly achieves the entire region. This is due to
the fact that in the low SNR regime only the best subcarrier is used. This is
realized with rate water-filling, even if not perfect but only ordinal
information, e.g. the ranking of the subcarriers, is present. The remaining
gap stems from the second term and the fact, that users can ''collide'', i.e.
have a common best subcarrier.

In contrast, in Fig. \ref{fig:dlc_high} the high SNR scenario is presented.
The bound improves as $s$ is increased up to $s=4$. From $s=5$ on, the bound
degrades once again. This is since it is not optimal to support the entire
rate only on one subcarrier, even if a user has only one best subcarrier.
Thus, the bound improves as $s$ is increased. Note, that for the case that
both users have similar rate requirements, i.e. the \emph{sum DLC case}, the
bound achieves a major part of the DLC and outperforms the time-sharing
strategy. The remaining gap on the axes is much bigger. However, the gap on
the axes can be reduced by sharing the subcarriers proportional to the users
rate requirements and using only the second term (thus making the conditioning
of the pdf needless). This is illustrated with the curve called \emph{prorated
FDMA}. The discontinuity stems from the switching of the subcarrier
allocation, since this is a discrete procedure. The dashed blue curve
indicates an \emph{achievable OFDMA DLC region}, which is given in Algorithm
\ref{ofdma_dlc_algorithm}: For any rate vector $\mathbf{R}$ and any channel
realization $\mathbf{h}$ the sum power minimization algorithm is usd to
calculate the optimal resource allocation. The resulting Lagrangian
multipliers $\boldsymbol{\mu}$ are taken to allocate the subcarriers according
to the maximum weighted channel rule $m_{k}=\arg\max_{m\in\mathcal{M}}\mu
_{m}h_{m,k}$. Once, the subcarrier allocation is done, the optimal resource
allocation is obtained by water-filling such that the rate requirements are met.

\begin{algorithm}[h]
\caption{OFDMA DLC Algorithm}\label{ofdma_dlc_algorithm}
\begin{algorithmic}
\STATE{\bf(1)} for given rate vector $\mathbf{R}
$ and channel realization $\mathbf{h}$
solve the minimum sum power problem with Alg \ref{alg:min_sum_power}
\STATE{\bf(2)} assign subcarriers according to $m_{k}=\arg\max_{m\in
\mathcal{M}}\mu_{m}h_{m,k}$
\FOR{$m=1$ to $M$}
\STATE{\bf(3)} do water-filling with respect to the rates
$R_{m,k}$ for user $m$ such that rate requirement $R_m$ is met
\ENDFOR
\end{algorithmic}
\end{algorithm}
This scheme requires perfect CSI but is computationally still relatively
simple due to the iterative water-filling principle. It can be seen from Fig.
\ref{fig:dlc_high} that the algorithm yields good results where any other FDMA
approach has to compete with.

\section{Conclusions}

\label{sec:conclusions}

We studied the delay limited capacity of OFDM systems. We have shown that
explicite expressions can be found in the low and high SNR regime even for the
challenging correlation structure of OFDM. Even though we presented our results
in the context of OFDM they are not restricted to this class but apply to
other channels such as MIMO as well. On the other hand, still a basic open
problem is the complete characterization of the DLC for all SNR and arbitrary
power delay profile. Here, we were not able to give universal bounds and it is
an interesting problem to show that the dependence is in general so-called
Schur-concave which implies that a uniform profile maximizes the DLC in all
cases. Furthermore, we analyzed the OFDM BC DLC region and derived lower
bounds based on rate water-filling. In the low SNR regime and concerning DLC
throughput, these bounds perform very well. To approach the DLC close to the
axes in the high SNR regime, a prorated strategy has to be used. All bounds
merely use order statistics and involve only ordinal -- and thus partial --
channel knowledge, which suggests savings for the design of future feedback
protocols. Moreover, an additional FDMA strategy using full channel state
information is proposed, performing very well over the entire region.

%
%
%
%
%
%
%
\begin{appendix}

\subsection{Proof of Theorem \ref{theorem:dist}}\label{proof_of_theorem:dist}

By Theorem 3 in \cite{wunder_06_inf} we have to
check that:

\begin{itemize}
\item $\mathbb{E}_{\mathbf{h}}\left(  \exp\left[  j\omega\Re e\left(
\sum_{l=1}^{L}\tilde{c}_{l}\,e^{-jl\theta_{k}}\right)  \right]  \right)
=e^{-\frac{\omega^{2}}{4}+\sum_{i=3}^{5}a_{i}\omega^{i}+O\left(  \omega
^{6}\right)  }$ holds for any $\theta_{k}:=2\pi k/K,k=0,\ldots,K-1$, and for
any real number $\omega$ in the non-empty interval $\left[  -d,d\right]  $ for
some $d>0$, and furthermore

\item $\frac{1}{K^{2}}\sum_{k_{1}=1}^{K}\sum_{k_{2}=1,k_{2}\neq k_{1}}%
^{K}\mathbb{E}_{\mathbf{h}}\left(  \exp\left[  j\Re e\left(  \sum_{l=1}%
^{L}\tilde{c}_{l}\left(  \omega_{1}e^{-jl\theta_{k_{1}-1}}+\omega
_{2}e^{-jl\theta_{k_{2}-1}}\right)  \right)  \right]  \right)  $

$=e^{-\frac{\omega_{1}^{2}}{4}-\frac{\omega_{2}^{2}}{4}+\sum_{i=3}^{5}%
a_{i}\left(  \left|  \omega_{1}\right|  +\left|  \omega_{2}\right|  \right)
^{i}+O\left(  \left(  \left|  \omega_{1}\right|  +\left|  \omega_{2}\right|
\right)  ^{6}\right)  }$ holds for all real numbers $\omega_{1},\omega_{2}$ in
$\left[  -d,d\right]  ^{2}$.
\end{itemize}

We show only the more complicated second condition. The first condition can be
easily deduced by our assumptions on the distributions and observing that the
subcarriers' real and imaginary parts are independent. We have
\begin{align*}
&  \mathbb{E}_{\mathbf{h}}\left(  \exp\left[  j\Re e\left(  \sum_{l=1}%
^{L}\tilde{c}_{l}\left(  \omega_{1}e^{-jl\theta_{k_{1}}}+\omega_{2}%
e^{-jl\theta_{k_{2}}}\right)  \right)  \right]  \right) \\
&  =e^{-\frac{1}{4L}\sum_{l=1}^{L}\left(  \omega_{1}\cos\left(  l\theta
_{k_{1}}\right)  +\omega_{2}\cos\left(  l\theta_{k_{2}}\right)  \right)
^{2}+\left(  \omega_{1}\sin\left(  l\theta_{k_{1}}\right)  +\omega_{2}%
\sin\left(  l\theta_{k_{2}}\right)  \right)  ^{2}}e^{O\left(  \left(  \left|
\omega_{1}\right|  +\left|  \omega_{2}\right|  \right)  ^{3}\right)  }.
\end{align*}
By analytic expansion of the first factor and observing that%
\begin{align*}
&  \frac{1}{4L}\sum_{k_{1}=0}^{K-1}\sum_{k_{2}=0,k_{2}\neq k_{1}}^{K-1}%
\sum_{l=1}^{L}\left(  \omega_{1}\cos\left(  l\theta_{k_{1}}\right)
+\omega_{2}\cos\left(  l\theta_{k_{2}}\right)  \right)  ^{2}+\left(
\omega_{1}\sin\left(  l\theta_{k_{1}}\right)  +\omega_{2}\sin\left(
l\theta_{k_{2}}\right)  \right)  ^{2}\\
&  =\frac{\omega_{1}^{2}}{4}+\frac{\omega_{2}^{2}}{4}+2\omega_{1}\omega
_{2}\sum_{k_{1}=0}^{K-1}\sum_{k_{2}=0,k_{2}\neq k_{1}}^{K-1}\sum_{l=1}%
^{L}\left(  \cos\left(  l\theta_{k_{1}}\right)  \cos\left(  l\theta_{k_{2}%
}\right)  +\sin\left(  l\theta_{k_{1}}\right)  \sin\left(  l\theta_{k_{2}%
}\right)  \right) \\
&  =\frac{\omega_{1}^{2}}{4}+\frac{\omega_{2}^{2}}{4}%
\end{align*}
where the last step is due to the standard trigonometric relation
\[
\sum_{k=0}^{K-1}\cos\left(  l\theta_{k}\right)  =\left\{
\begin{array}
[c]{cc}%
K & l=0\\
0 & l\neq0,l<K
\end{array}
\right.
\]
we have finally%
\begin{align*}
&  e^{\left(  -\frac{1}{4L}\sum_{l=1}^{L}\left(  \omega_{1}\cos\left(
l\theta_{k_{1}}\right)  +\omega_{2}\cos\left(  l\theta_{k_{2}}\right)
\right)  ^{2}+\left(  \omega_{1}\sin\left(  l\theta_{k_{1}}\right)
+\omega_{2}\sin\left(  l\theta_{k_{2}}\right)  \right)  ^{2}+O\left(  \left(
\left|  \omega_{1}\right|  +\left|  \omega_{2}\right|  \right)  ^{3}\right)
\right)  }\\
&  =e^{-\frac{\omega_{1}^{2}}{4}-\frac{\omega_{2}^{2}}{4}+\sum_{i=3}^{5}%
a_{i}\left(  \left|  \omega_{1}\right|  +\left|  \omega_{2}\right|  \right)
^{i}+O\left(  \left(  \left|  \omega_{1}\right|  +\left|  \omega_{2}\right|
\right)  ^{6}\right)  }%
\end{align*}
which is the desired result. The proof that non-uniform PDP can not improve
this bound follows from union bound and is omitted \cite{wunder_06_inf}.

\subsection{Proof of Lemma \ref{lemma:gauss}}\label{proof_of_lemma:gauss}

The order of the lower and upper bound was already
derived in \cite{wunder_06_inf}. Due to the correlation structure imposed by
the FFT and Rayleigh fading with uniform PDP the channel gains $h_{k_{1}%
+\left(  a-1\right)  k_{2}},k_{2}\in\left\{  1,\ldots,L\right\}  $, are
independent for any $k_{1}\in\left\{  1,\ldots,a\right\}  $ for some
$a\in\mathbb{N}$ where $a=K/L$. Since the maximum $\left\|  \left(  h_{k_{1}%
},h_{k_{1}+a},\ldots,h_{k_{1}+a\left(  L-1\right)  }\right)  \right\|
_{\infty}$ is below or equal $x$ if $h_{\infty}$ is below or equal $x$ but not
conversely, it follows the lower bound \cite{wunder_06_inf}%
\[
\Pr\left(  h_{\infty}\leq x\right)  \leq e^{-L^{\varepsilon}}%
\]
which is the desired lower bound if we set $\varepsilon=\gamma\log\left[
\log\left(  L\right)  \right]  /\log\left(  L\right)  $ (in fact, obviously,
it is even stronger and can be to strengthened to fall off with $L^{-1}$
instead of order $\log^{-1}\left(  L\right)  $). The upper bound is obtained
by observing that for any $a\geq1$ we have by the FFT structure%
\[
\Pr\left(  h_{\infty}>x\right)  \leq a\,\left[  1-\left(  1-e^{-x}\right)
^{L}\right]  .
\]
Setting this time $x=\left(  1+\varepsilon\right)  \log\left(  L\right)  $ for
some $\varepsilon>0$ yields
\begin{align*}
\Pr\left(  h_{\infty}>\left(  1+\varepsilon\right)  \log\left(  L\right)
\right)   &  \leq a\,\left[  1-\left(  1-\frac{1}{L^{\left(  1+\varepsilon
\right)  }}\right)  ^{L}\right] \\
&  =a\,\left(  1-\exp\left[  L\log\left(  1-\frac{1}{L^{\left(  1+\varepsilon
\right)  }}\right)  \right]  \right) \\
&  \leq a\,\left(  1-\exp\left[  -\frac{L^{-\varepsilon}}{1-L^{-\left(
1+\varepsilon\right)  }}\right]  \right) \\
&  \leq\frac{a\,L^{-\varepsilon}}{1-L^{-\left(  1+\varepsilon\right)  }}\\
&  \leq\frac{aL}{L-L^{-\varepsilon}}L^{-\varepsilon}%
\end{align*}
using $\log\left(  1-x\right)  \geq-\frac{x}{1-x}$ and $e^{-x}\geq1-x$. Hence,
if we set $\varepsilon=\frac{\gamma\log\left[  \log\left(  L\right)  \right]
}{\log\left(  L\right)  }$ we have
\[
\Pr\left(  h_{\infty}>\log\left(  L\right)  +\log\left[  \log\left(  L\right)
\right]  \right)  \leq\frac{\kappa}{\log^{\gamma}\left(  L\right)  }.
\]
Combining this with the stronger lower bound yields the result.

\subsection{Proof of Proposition \ref{prop:high3}}\label{proof_of_prop:high3}

We can frequently use the following well-known
inequality: suppose that $f_{1},\ldots,f_{K}$ are functions defined on a
domain $\Omega$ equipped with some measure $F$ with $f_{k}\in\mathcal{L}%
^{p_{k}}\left(  \Omega\right)  $ and $\sum_{i=1}^{K}p_{k}^{-1}=1$ then
\cite{LieLos:1997}:%
\begin{equation}
\left|  \int_{\Omega}\prod_{k=1}^{K}f_{k}\left(  x\right)  \,dF\left(
x\right)  \right|  \leq\prod_{k=1}^{K}\left(  \int_{\mathbb{R}_{+}}f^{p_{k}%
}\left(  x\right)  \,dF\left(  x\right)  \right)  ^{1/p_{k}} \label{eqn:lieb}%
\end{equation}
This inequality is tailored for the situation at hand: suppose that for some
$a,b>0$
\begin{align*}
\mathbb{E}_{\mathbf{h}}\left(  \overline{h}\right)   &  =\int_{\mathbb{R}%
_{+}^{K}}\prod_{k=1}^{K}h^{-\frac{1}{K}}\,dF_{\mathbf{h}}\left(
\mathbf{h}\right) \\
&  =\int_{\mathbb{R}_{+}^{K}}\prod_{k_{1}=1}^{a}\prod_{k_{2}=1}^{b}%
h_{k_{1}+\left(  k_{2}-1\right)  a}^{-\frac{1}{K}}\,dF_{\mathbf{h}}\left(
\mathbf{h}\right)
\end{align*}
which yields by application of (\ref{eqn:lieb}):%
\begin{equation}
\mathbb{E}_{\mathbf{h}}\left(  \overline{h}\right)  \leq\prod_{k_{1}=1}%
^{a}\left[  \int_{\mathbb{R}_{+}^{K}}\prod_{k_{2}=1}^{b}h_{k_{1}+\left(
k_{2}-1\right)  a}^{-\frac{1}{b}}\,dF_{\mathbf{h}}\left(  \mathbf{h}\right)
\right]  ^{\frac{1}{a}} \label{eqn:e_h}%
\end{equation}
The inner term on the RHS of (\ref{eqn:e_h}) generally means multidimensional
integration with usually dependent random variables which cannot be directly
carried out. Hence, we resort to some bounding techniques and have to show
that an upper bound holds for the inner integral independent of $k_{1}$.

In order to obtain an upper bound on the inner term on the RHS of
(\ref{eqn:e_h}) choose some $\epsilon>0$ and subdivide the integration domain
in parts where in each dimension the range of integration is either in the
interval $\left[  0,\epsilon\right]  $ or outside this interval. For those
dimensions that are within this interval we bound the corresponding marginal
density by the constant $c_{s}$ and calculate the remaining integral while for
those dimensions that are outside the interval we simply set the values of the
integrand to $\epsilon$. Suppose that $l$ dimensions are within the interval.
Since it does not matter what particular dimensions are chosen for this
decomposition we have $l$ out of $b$ possibilities that can be equally
treated. Hence, we can write
\begin{align*}
\mathbb{E}_{\mathbf{h}}\left(  \overline{h}\right)   &  \leq\int
_{\mathbb{R}_{+}^{K}}\prod_{k=1}^{b}h_{k_{1}+\left(  k_{2}-1\right)
a}^{-\frac{1}{b}}\,dF_{\mathbf{h}}\left(  \mathbf{h}\right) \\
&  \leq c_{s}\sum_{l=0}^{b-1}\binom{b}{l}\prod_{k=1}^{l}\epsilon^{-\frac{1}%
{b}}\int_{\left[  0,\epsilon\right]  ^{b-l}}\prod_{k=1}^{b-l}h_{k_{1}+\left(
k_{2}-1\right)  a}^{-\frac{1}{b}}\,dh+\frac{1}{\epsilon}\\
&  =c_{s}\sum_{l=0}^{b}\binom{b}{l}\left(  \frac{b}{b-1}\right)
^{b-l}\epsilon^{b-l-1}+\frac{1-c_{s}}{\epsilon},
\end{align*}
and since
\[
\sum_{l=0}^{b}\binom{b}{l}\left(  1-\epsilon\right)  ^{l}\epsilon^{b-l}=1
\]
for any $0\leq\epsilon\leq1$ it follows
\[
\mathbb{E}_{\mathbf{h}}\left(  \overline{h}\right)  \leq c_{s}b\left(
\frac{b}{b-1}\right)  ^{b}+\left(  1-c_{s}\right)  b
\]
where $\epsilon=1/b<1$.

\subsection{Proof of Proposition \ref{prop:high4}}\label{proof_of_prop:high4}

We can nicely use H\"{o}lder's inequality. Swapping
expectation and product operator we have by partial integration
\begin{align*}
\int_{\mathbb{R}_{+}}h^{-\frac{1}{b}}\,f\left(  h\right)  \,dh  &
=\underbrace{\left.  h^{-\frac{1}{b}+1}f\left(  h\right)  \right|
_{0}^{\infty}}_{=0}-\frac{b}{b-1}\int_{\mathbb{R}_{+}}h^{-\frac{1}{b}%
+1}\,f^{\prime}\left(  h\right)  \,dh\\
&  =-\frac{b}{b-1}\int_{\mathbb{R}_{+}}h^{-\frac{1}{b}+1}\,f^{\prime}\left(
h\right)  \,dh
\end{align*}
since $f$ finite everywhere. Define $\mathbb{R}_{+}^{\circ}:=\left\{
h:f^{\prime}\left(  h\right)  <0\right\}  $ and let $0\leq a_{i}^{-}<b_{i}%
^{-}\leq\infty,i\in\mathbb{N}$, be the interval boundaries of $h$ where
$f^{\prime}\left(  h\right)  \leq0$ as well as $0\leq a_{i}^{+}<b_{i}^{+}%
\leq\infty,i\in\mathbb{N}$, those where $f^{\prime}\left(  h\right)  \geq0$.
Representing $\left(  -f^{\prime}\right)  $ as $\left(  -f^{\prime}\right)
=\left(  -f^{\prime}\right)  ^{\frac{b-1}{b}}\circ\left(  -f^{\prime}\right)
^{\frac{1}{b}}$ in $\mathbb{R}_{+}^{\circ}$ yields%
\[
-\frac{b}{b-1}\int_{\mathbb{R}_{+}^{\circ}}h^{-\frac{1}{b}+1}\,f^{\prime
}\left(  h\right)  \,dh=\frac{b}{b-1}\int_{\mathbb{R}_{+}^{\circ}}%
h^{\frac{b-1}{b}}\left(  -f^{\prime}\left(  h\right)  \right)  ^{\frac{b-1}%
{b}}\left(  -f^{\prime}\left(  h\right)  \right)  ^{\frac{1}{b}}\left(
h\right)  \,dh
\]
Setting $p=b/\left(  b-1\right)  $ and $q=b$%
\begin{align*}
\int_{\mathbb{R}_{+}^{\circ}}h^{-\frac{1}{b}}\,\left(  -f^{\prime}\left(
h\right)  \right)  \,dh  &  \leq\left(  \int_{\mathbb{R}_{+}^{\circ}}\left(
-f^{\prime}\left(  h\right)  \right)  \,dh\right)  ^{\frac{1}{b}}\left(
\int_{\mathbb{R}_{+}^{\circ}}h\left(  -\,f^{\prime}\left(  h\right)  \right)
\,dh\right)  ^{\frac{b-1}{b}}\\
&  \leq\left(  \sum_{i\geq1}f\left(  a_{i}^{-}\right)  -f\left(  b_{i}%
^{-}\right)  \right)  ^{\frac{1}{b}}\left(  \int_{\mathbb{R}_{+}^{\circ}%
}h\,\left(  -f^{\prime}\left(  h\right)  \right)  \,dh\right)  ^{\frac{b-1}%
{b}}%
\end{align*}
Again by partial integration for the last term
\begin{align*}
-\int_{\mathbb{R}_{+}^{\circ}}h\,f^{\prime}\left(  h\right)  \,dh  &
=-\left.  hf\left(  h\right)  \right|  _{0}^{\infty}+\int_{\mathbb{R}%
_{+}^{\circ}}f\left(  h\right)  \,dh\\
&  \leq1
\end{align*}
and since
\begin{align*}
-\frac{b}{b-1}\int_{\left(  \mathbb{R}_{+}^{\circ}\right)  ^{C}}h^{-\frac
{1}{b}+1}\,f^{\prime}\left(  h\right)  \,dh  &  \leq-\frac{b}{b-1}\sum
_{i\geq1}a_{i}^{\frac{b-1}{b}}\int_{\left[  a_{i}^{+},b_{i}^{+}\right]
}f^{\prime}\left(  h\right)  \,dh\\
&  =\,-\frac{b}{b-1}\sum_{i\geq1}a_{i}^{\frac{b-1}{b}}\left(  f\left(
b_{i}^{+}\right)  -f\left(  a_{i}^{+}\right)  \right)
\end{align*}
we have finally
\begin{align*}
&  \left(  \int_{\mathbb{R}_{+}}h^{-\frac{1}{b}}\,f\left(  h\right)
\,dh\right)  ^{b}\\
&  \leq\left(  \frac{b}{b-1}\right)  ^{b}\left[  \left(  \sum_{i\geq1}f\left(
a_{i}^{-}\right)  -f\left(  b_{i}^{-}\right)  \right)  ^{1/b}-\sum_{i\geq
1}a_{i}^{\frac{b-1}{b}}\left(  f\left(  b_{i}^{+}\right)  -f\left(  a_{i}%
^{+}\right)  \right)  \right]  ^{b}%
\end{align*}
which concludes the proof.

\subsection{Proof of Proposition \ref{prop:high5}}\label{proof_of_prop:high5}

Suppose $K,L$ to be even integers. We can apply
inequality (\ref{eqn:e_h}) with $a=K/2$ and $b=2$:%
\[
\int_{\mathbb{R}_{+}^{K}}\prod_{k=1}^{K}h_{k}^{-\frac{1}{K}}\,dF_{\mathbf{h}%
}\left(  \mathbf{h}\right)  \leq\prod_{l=1}^{K/2}\left(  \int_{\mathbb{R}%
_{+}^{K}}h_{l}^{-1/2}h_{l+K/2}^{-1/2}\,dF_{\mathbf{h}}\left(  \mathbf{h}%
\right)  \right)  ^{2/K}%
\]
Since $h_{k}=|\mathbf{\tilde{c}}^{T}\boldsymbol{\zeta}_{k}|^{2}$ with
$\boldsymbol{\zeta}_{k}:=[1,e^{-2\pi j\left(  k-1\right)  /K},...,e^{-2\pi
j\left(  k-1\right)  \left(  K-1\right)  /K}]^{T}$ we can write for the inner
integral%
\begin{equation}
\int_{\mathbb{R}_{+}^{K}}h_{l}^{-1/2}h_{l+K/2}^{-1/2}\,dF_{\mathbf{h}}\left(
\mathbf{h}\right)  =\iint_{\mathbb{R}^{2L}}\prod_{k=1}^{2}\left|
\mathbf{\tilde{c}}^{T}\boldsymbol{\zeta}_{\left(  k-1\right)  K/2+l}\right|
^{-1}\prod_{k=1}^{L}f_{\tilde{c}_{k}^{r}}^{r}\left(  \tilde{c}_{k}^{r}\right)
f_{\tilde{c}_{k}^{i}}\left(  \tilde{c}_{k}^{i}\right)  \,d\mathbf{\tilde{c}%
}^{r}d\mathbf{\tilde{c}}^{i} \label{eqn:trans}%
\end{equation}
where $f^{r},f^{i}$ are the bounded densities of real and imaginary parts of
the complex path gains and $\left(  \cdot\right)  ^{r,i}$ is a shorthand
notation for real and imaginary part operators. Let $l\in\left[  1,K/2\right]
$ be arbitrary but fixed. The following change of coordinates is based on the
observation that for $K,L$ even $\boldsymbol
{\zeta}_{l},\boldsymbol{\zeta}_{l+K/2}$ are orthogonal in $\mathbb{C}%
^{L},\forall l$. For $l=1$ this is obvious since the first and $K/2$-th vector
consist of an even number of binary $\pm1$'s only. For $l>1$ the same follows
from the fact that two orthogonal vectors remain orthogonal if the are both
multiplied by the same complex phase factors.

Hence, there exist $\boldsymbol{\zeta}_{i}^{z},i=2,\ldots L-2$, that can be
chosen to span the basis of the orthogonal complement. By change of
coordinates $\mathbf{\tilde{c}\rightarrow}\left(  \mathbf{\tilde{h}}%
^{z}\right)  ,\tilde{h}_{1}^{z}=\mathbf{\tilde{c}}^{T}\boldsymbol{\zeta}%
_{l},\tilde{h}_{2}^{z}=\mathbf{\tilde{c}}^{T}\boldsymbol{\zeta}_{K/2+l}%
,\tilde{h}_{i}^{z}=\mathbf{\tilde{c}}^{T}\boldsymbol{\zeta}_{i}^{z},i=2,\ldots
L-2,$ where $\boldsymbol{\zeta}_{l},\boldsymbol{\zeta}_{l+K/2},\boldsymbol
{\zeta}_{i}^{z},\forall i$ is an orthogonal transformation and the Jacobian
equals $1/L^{L}$ and, further, $\tilde{h}_{k}^{z}\rightarrow(|\tilde{h}%
_{k}^{z}|^{2},\varphi_{k}^{z})=\left(  R_{k}^{z},\varphi_{k}^{z}\right)
,k=1,\ldots L,$ of which the Jacobian is $1/2$ $\forall k$ we obtain by
assuming $f_{\tilde{c}_{k}^{r}}\left(  x\right)  \leq v_{k}^{1/2}%
e^{-\alpha\left|  x\right|  ^{2}}$ and $f_{\tilde{c}_{k}^{i}}\left(  x\right)
\leq v_{k}^{1/2}e^{-\alpha\left|  x\right|  ^{2}},\forall k$:
\begin{align*}
\text{RHS of (\ref{eqn:trans})}  &  \leq\frac{\pi^{L}}{L^{L}}\int
_{\mathbb{R}_{+}^{L}}\left(  R_{1}^{z}\right)  ^{-\frac{1}{2}}\left(
R_{2}^{z}\right)  ^{-\frac{1}{2}}\quad...\\
&  \prod_{k=1}^{L}v_{k}^{1/2}e^{-\alpha\left|  \left(  \frac{1}{L}%
\boldsymbol{\zeta}_{\left(  k-1\right)  K/L+l}^{H}\mathbf{\tilde{h}}%
^{z}\left(  \mathbf{R}^{z},\boldsymbol{\phi}^{z}\right)  \right)  ^{r}\right|
^{2}}v_{k}^{1/2}e^{-\alpha\left(  \frac{1}{L}\boldsymbol{\zeta}_{\left(
k-1\right)  K/L+l}^{H}\mathbf{\tilde{h}}^{z}\left(  \left(  \mathbf{R}%
^{z},\boldsymbol{\phi}^{z}\right)  \right)  \right)  ^{i}}\,d\boldsymbol{\phi
}^{z}d\mathbf{R}^{z}\\
&  =\frac{\pi^{L}\prod_{k=1}^{L}v_{k}}{L^{L}}\prod_{k=1}^{2}\int
_{\mathbb{R}_{+}}\left(  R_{k}^{z}\right)  ^{-\frac{1}{2}}\,e^{-\frac{\alpha
R_{k}^{z}}{L}}\,dR^{z}\prod_{k=3}^{L}\int_{\mathbb{R}_{+}}e^{-\frac{\alpha
R_{k}^{z}}{L}}\,dR^{z}\\
&  \leq\frac{\pi^{L}\prod_{k=1}^{L}v_{k}}{L^{L}}\left(  \frac{2}{2-1}\right)
^{2}\left(  \frac{L}{\alpha}\right)  \left(  \frac{L}{\alpha}\right)
^{L-2}\quad\text{(apply Prop. \ref{prop:high4})}\\
&  =\frac{4\pi^{L}\prod_{k=1}^{L}v_{k}}{L^{L}}\left(  \frac{L}{\alpha}\right)
^{L-1}%
\end{align*}

\begin{remark}
For many fading distributions the claim $f_{\tilde{c}_{k}^{r,i}}\left(
x\right)  \leq v_{k}^{1/2}e^{-\alpha\left|  x\right|  ^{2}}$ might be too
restrictive and shall be replaced by $f_{\tilde{c}_{k}^{r,i}}\left(  x\right)
\leq\max\left\{  c^{\text{(max)}},v_{k}^{1/2}e^{-\alpha\left|  x\right|  ^{2}%
}\right\}  $ where $c^{\text{(max)}}>0$ is some global constant. The latter
inequality, however, does not separate over $\mathbb{R}^{L}$ as required in
the proof here. In this situation, we can obtain better bounds for specific
cases by combining the techniques of Prop. \ref{prop:high3} - Prop.
\ref{prop:high5}, e.g. by splitting up the integration domain similar to
Prop. \ref{prop:high3}.
\end{remark}

\end{appendix}

%
%
%
%
%
%
%
%
%
%
%
%
%
%
%
%
%
%
%
%
%
%
%
%
%
%
%
%
%
\bibliographystyle{ieeebib}
\bibliography{mybibliography,../../../habil/litermod}
\newpage\begin{figure}[ptbptbptbptbptbptb]
\begin{center}
\includegraphics[width=10cm]{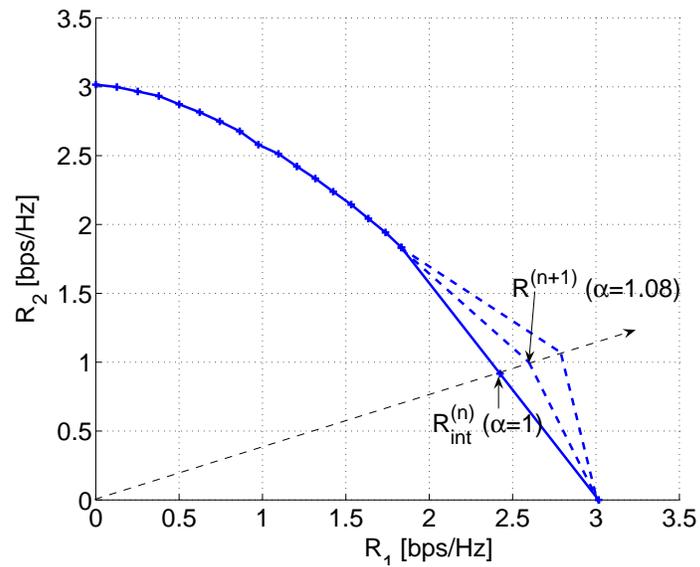}
\end{center}
\caption{Example for an iteration of the described algorithm to calculate the
OFDM DLC region. OFDM MAC DLC region for 2 users with 7 i.i.d taps each and 16
subcarriers at $10dB$ }%
\label{fig:dlc}%
\end{figure}
\end{document}